\documentclass[letterpaper,twocolumn,10pt]{article}
\usepackage{arxiv_sty}

\usepackage{filecontents}

\usepackage[utf8]{inputenc} 
\usepackage[T1]{fontenc}    
\usepackage{url}            
\usepackage{booktabs}       
\usepackage{amsfonts}       
\usepackage{nicefrac}       
\usepackage{microtype}      
\usepackage{xcolor}         
\usepackage{wrapfig}
\usepackage{multirow}
\usepackage{amsmath}

\usepackage[capitalize]{cleveref}
\crefname{section}{Sec.}{Secs.}
\Crefname{section}{Section}{Sections}
\Crefname{table}{Table}{Tables}
\crefname{table}{Tab.}{Tabs.}

\usepackage[amsmath, thmmarks]{ntheorem}
\theoremseparator{:}

\theoremsymbol{\ensuremath{\square}}

\usepackage[pdftex]{graphicx}
\usepackage{subcaption}

\usepackage[normalem]{ulem}
\usepackage{caption}
\usepackage{mathtools}
\usepackage{tikz}
\usepackage{xspace}
\usepackage{textcomp}
\usepackage{rotating}

\newcommand{\zb}[1]{\textbf{#1}}

\begin{document}

\date{}

\title{\Large \bf How Does a Deep Learning Model Architecture Impact Its Privacy? \\ A Comprehensive Study of Privacy Attacks on CNNs and Transformers}

\author{%
{\rm Guangsheng Zhang\textsuperscript{1}}\ \ \ 
{\rm Bo Liu\textsuperscript{1}}\ \ \
{\rm Huan Tian\textsuperscript{1}}\ \ \
{\rm Tianqing Zhu\textsuperscript{1}}\ \ \
\\
{\rm Ming Ding\textsuperscript{2}}\ \ \
{\rm Wanlei Zhou\textsuperscript{3}}\ \ \
\\
\\
\textsuperscript{1}\textit{University of Technology Sydney}\ \ \ 
\textsuperscript{2}\textit{Data 61, Australia}\ \ \ 
\textsuperscript{3}\textit{City University of Macau}
}

\maketitle

\begin{abstract}
  As a booming research area in the past decade, deep learning technologies have been driven by big data collected and processed on an unprecedented scale. 
  However, privacy concerns arise due to the potential leakage of sensitive information from the training data.
  Recent research has revealed that deep learning models are vulnerable to various privacy attacks, 
  including membership inference attacks, 
  attribute inference attacks, 
  and gradient inversion attacks.
  Notably, the efficacy of these attacks varies from model to model.
  In this paper, 
  we answer a fundamental question: \textit{Does model architecture affect model privacy?}
  By investigating representative model architectures from convolutional neural networks (CNNs) to Transformers,
  we demonstrate that Transformers generally exhibit higher vulnerability to privacy attacks than CNNs.
  Additionally, we identify the micro design of activation layers, 
  stem layers, 
  and LN layers, 
  as major factors contributing to the resilience of CNNs against privacy attacks, while the presence of attention modules is another main factor that exacerbates the privacy vulnerability of Transformers.
  Our discovery reveals valuable insights for deep learning models to defend against privacy attacks and inspires the research community to develop privacy-friendly model architectures.
\end{abstract}

\section{Introduction}
\label{sec:introduction}

Deep learning has been gaining massive attention over the past several years.
Training deep learning models requires collecting and processing user data, 
which raises significant privacy concerns.
The data gathered during the training phase often contains sensitive information that malicious parties can access or retrieve. 
Various privacy attacks targeting deep learning models have demonstrated this vulnerability extensively.
One prominent type of attack is membership inference, 
which focuses on determining whether a specific data sample belongs to the training data~\cite{shokriMembershipInferenceAttacks2017, salemMLLeaksModelData2019}. 
Another attack is attribute inference, 
which aims to uncover implicit attributes learned by the model beyond the intended target attribute~\cite{melisExploitingUnintendedFeature2019, songOverlearningRevealsSensitive2020}. 
Additionally, gradient inversion attacks pose a significant threat by attempting to reconstruct the information of the training data from the gradients of the model~\cite{fredriksonModelInversionAttacks2015, geipingInvertingGradientsHow2020}. 
These attacks empower adversaries to exploit deep learning models for extracting sensitive data.

Prior research has established that overfitting is one of the primary causes of privacy leakage in deep learning models~\cite{shokriMembershipInferenceAttacks2017,heQuantifyingMitigatingPrivacy2021,chenWhenMachineUnlearning2021,liuWhenMachineLearning2021}. 
In general, 
overfitting occurs when models excessively learn specific details from the training data, 
which can lead to inadvertent privacy breaches. 
Surprisingly, we discover that even when models exhibit comparable levels of overfitting, 
the effectiveness of attacks varies across different models. 
This observation raises intriguing questions as to why certain deep learning models are more susceptible to privacy attacks than others, 
a puzzle that researchers have not fully comprehended.
Consequently, we conjecture that other factors beyond overfitting might also contribute to the increased vulnerability of some deep learning models to privacy attacks. 
Though existing literature has explored model robustness and explainability~\cite{baiAreTransformersMore2021,raghuVisionTransformersSee2021}, 
the privacy leakage of the model architectures remains underexplored.
Therefore, we are motivated to address this critical gap by answering the following question: 
\textit{How does a model's architecture affect its privacy preservation capability?}

In this paper, 
we approach this question by comprehensively analyzing different deep learning models under various state-of-the-art privacy attacks. 
Our investigation focuses on two widely adopted deep learning model architectures: convolutional neural networks (CNNs) and Transformers.
CNN-based models have been dominant in computer vision, 
thanks to its sliding-window strategy, which extracts local information from images effectively. 
Transformers, 
initially introduced in natural language processing (NLP), 
have gained popularity in computer vision by capturing large receptive fields through attention mechanisms, 
resulting in comparable accuracy performance against CNNs.
The tremendous achievements and wide usage of these two model architectures provide an excellent opportunity for us to make a comparative analysis regarding model privacy risks.
Through our investigation, 
we make an intriguing discovery: \textit{Transformers, in general, exhibit higher vulnerability to mainstream privacy attacks than CNNs.}

While Transformers and CNNs have different designs in many aspects, 
we investigate whether some key modules in the model architecture have a major impact on privacy risks. 
To this end, 
we evaluate the privacy leakage of several major modules in a Transformer architecture by sending only selected gradients to the gradient inversion attacks and discover that attention modules cause significant privacy leakage.
Moreover, 
we start with a popular CNN-based model, 
ResNet-50~\cite{heDeepResidualLearning2016}, 
and gradually morph the model to incorporate the key designs of Transformers. 
This leads us to the structure of ConvNeXt~\cite{liuConvNet2020s2022}. 
We evaluate the privacy leakage through this process and identify several key components that have a significant impact on privacy risks: 
(1) the design of the activation layers;
(2) the design of stem layers;
(3) the design of LN layers. 
We further conduct ablation studies to verify our discoveries and propose solutions to mitigate the privacy risks.

In summary, 
our contributions in this paper are summarized as follows:
\begin{itemize}
    \itemsep-0.28em 
    \item For the first time, 
    we investigate the impact of model architectures and micro designs on privacy risks.
    \item We evaluate the privacy vulnerabilities of two widely adopted model architectures, i.e., CNNs and Transformers, using three prominent privacy attack methods: (1) membership inference attacks, (2) attribute inference attacks, and (3) gradient inversion attacks. Our analysis reveals that Transformers exhibit higher vulnerabilities to these privacy attacks than CNNs.
    \item We identify three key factors: 
    (1) the design of activation layers, 
    (2) the design of stem layers, 
    and (3) the design of LN layers, 
    that significantly contribute to the enhanced resilience of CNNs in comparison to Transformers. 
    We also discover that the presence of attention modules in Transformers could make them susceptible to privacy attacks.
    \item We propose solutions to mitigate the vulnerabilities of model architectures: modifying model components and adding perturbations as defense mechanisms.
\end{itemize}

\section{Related Work}
\label{sec:related_work}

\subsection{CNNs and Vision Transformers}

\textbf{Convolutional Neural Networks (CNNs)} are a type of neural network that employs convolutional layers to extract features from input data.
In contrast to fully connected networks, 
CNNs use convolutional kernels to connect small samples to neurons for feature extraction, reducing the number of model parameters and enabling the recognition of local features.
Various techniques are employed to construct a CNN model, including padding, pooling, dilated convolution, group convolution, and more.

The concept of convolutional neural networks (CNNs) dates back to the 1980s~\cite{lecunBackpropagationAppliedHandwritten1989}.
However, the invention of AlexNet~\cite{krizhevskyImageNetClassificationDeep2012} makes CNNs the most prominent models in computer vision.
Subsequent research improved the accuracy and efficiency of models~\cite{simonyanVeryDeepConvolutional2015,szegedyGoingDeeperConvolutions2015}.
ResNet~\cite{heDeepResidualLearning2016} addressed the challenge of training deep networks using skip connections.
Other notable networks consist of Inception~\cite{szegedyRethinkingInceptionArchitecture2016}, MobileNet~\cite{howardSearchingMobileNetV32019}, ResNeXt~\cite{xieAggregatedResidualTransformations2017}, EfficientNet~\cite{tanEfficientNet2019}, RegNet~\cite{radosavovicDesigningNetworkDesign2020}, ConvNeXt~\cite{liuConvNet2020s2022}.

\textbf{Vision Transformers}, originating from natural language processing, divide the input image into multiple patches, forming a one-dimensional sequence of token embeddings.
Their exceptional performance can be attributed to the multi-head self-attention modules~\cite{vaswaniAttentionAllYou2017}.
The attention mechanism has significantly contributed to the advancement of natural language processing~\cite{devlinBERTPretrainingDeep2019,yangXLNetGeneralizedAutoregressive2019,brownLanguageModelsAre2020}, subsequently leading to the introduction of Transformers in the field of computer vision as Vision Transformers (ViT)~\cite{dosovitskiyImageWorth16x162021}.
Research has shown that ViTs can surpass CNNs in various downstream tasks~\cite{dosovitskiyImageWorth16x162021,steinerHowTrainYour2022}.
Later Transfomer models have focused on numerous improvements of ViTs,
such as Tokens-to-Token ViT~\cite{yuanTokenstoTokenViTTraining2021}, Swin Transformers~\cite{liuSwinTransformerHierarchical2021}, DeiT~\cite{touvronTrainingDataefficientImage2021}, MViT~\cite{liMViTv2ImprovedMultiscale2022}, DaViT~\cite{dingDaViTDualAttention2022}.

Numerous studies have compared CNNs and Transformers from the perspectives of robustness~\cite{baiAreTransformersMore2021,paulVisionTransformersAre2022,wangCanCNNsBe2023} and explainability~\cite{raghuVisionTransformersSee2021}. 
However, our research diverges from previous works by concentrating on the privacy leakage inherent in both CNNs and Transformers.

\subsection{Privacy Attacks on Deep Learning Models}

A primary concern in deep learning privacy is that the model may reveal sensitive information from the training dataset. 
An adversary can exploit various approaches to compromise privacy, including predicting whether a particular sample is in the model's training dataset via membership inference attacks,
or disclosing the implicit attributes of data samples via attribute inference attacks,
or even recovering private data samples utilized in training a neural network through gradient inversion attacks.

\textbf{Membership inference attacks} were initially introduced in~\cite{shokriMembershipInferenceAttacks2017}, where an attack model was employed to distinguish member samples from non-member samples in the training data.
To execute these attacks, shadow models would mimic the behavior of victim models~\cite{shokriMembershipInferenceAttacks2017,salemMLLeaksModelData2019}.
Prediction results from victim models were gathered for attack model training.
Usually, the confidence scores or losses were utilized~\cite{shokriMembershipInferenceAttacks2017}, 
but more recent work (label-only attacks) applied prediction labels to launch attacks successfully~\cite{choquette-chooLabelOnlyMembershipInference2021,liMembershipLeakageLabelOnly2021}.
The attacks could also be executed by designing a metric with a threshold by querying the shadow model~\cite{songSystematicEvaluationPrivacy2021}.
Some researchers expanded the attacks into new domains, 
including generative models~\cite{hayesLOGANMembershipInference2019,chenGANLeaksTaxonomyMembership2020}, 
semantic segmentation~\cite{heSegmentationsLeakMembershipInference2020,zhangLabelOnlyMembershipInference2022}, 
federated learning~\cite{nasrComprehensivePrivacyAnalysis2019,truexDemystifyingMembershipInference2019}, 
and transfer learning~\cite{songInformationLeakageEmbedding2020,zouPrivacyAnalysisDeep2020}.
Other researchers relaxed the attack assumptions and improved the attacks, including discussion on white-box/black-box access for the attacks~\cite{sablayrollesWhiteboxVsBlackbox2019}, providing more metrics (i.e. ROC curves and the true positive rate at a low false positive rate) to measure the attack performance more accurately~\cite{longPragmaticApproachMembership2020,jayaramanRevisitingMembershipInference2021,carliniMembershipInferenceAttacks2022,yeEnhancedMembershipInference2022,watsonImportanceDifficultyCalibration2022}.
We select~\cite{shokriMembershipInferenceAttacks2017,salemMLLeaksModelData2019,carliniMembershipInferenceAttacks2022} as our baseline methods.

\textbf{Attribute inference attacks}, another significant category of privacy attack methods, attempt to reveal a specific sensitive attribute of a data sample by analyzing the posteriors of the victim model trained by the victim dataset.
Some early research launched the attacks by generating input samples with different sensitive attributes and observed the victim model output~\cite{fredriksonPrivacyPharmacogeneticsEndtoEnd2014,yeomPrivacyRiskMachine2018}.
However, these methods could only work in structured data.
Later research improved the attacks with victim model representations~\cite{melisExploitingUnintendedFeature2019,songOverlearningRevealsSensitive2020}.
They also claimed that the overlearning feature of deep learning models caused the execution of the attacks~\cite{songOverlearningRevealsSensitive2020}.
Attributes could also be inferred through a relaxed notion~\cite{zhaoFeasibilityAttributeInference2021}, model explanations~\cite{dudduInferringSensitiveAttributes2022}, label-only settings~\cite{mehnazAreYourSensitive2022}, or imputation analysis~\cite{jayaramanAreAttributeInference2022}.
As we aim to infer attributes from visual data, we select~\cite{melisExploitingUnintendedFeature2019,songOverlearningRevealsSensitive2020} as baseline methods.

\textbf{Gradient inversion attacks} primarily aim to reconstruct training samples at the local clients in federated learning.
Using the publicly shared gradients in the server, 
adversaries can execute the attacks by reconstructing the training samples using gradient matching.
DLG~\cite{zhuDeepLeakageGradients2019} and its variant, iDLG ~\cite{zhaoIDLGImprovedDeep2020}, were the early attacks to employ an optimization-based technique to reconstruct the training samples.
Later research like Inverting Gradients~\cite{geipingInvertingGradientsHow2020} and GradInversion~\cite{yinSeeGradientsImage2021} improved the attack performance by incorporating regularizations into the optimization process.
APRIL~\cite{luAPRILFindingAchilles2022} and GradViT~\cite{hatamizadehGradViTGradientInversion2022} further developed the attack methods to extract sensitive information from Transformers.
The use of Generative Adversarial Networks (GANs) in some gradient inversion attack methods~\cite{liAuditingPrivacyDefenses2022} can have a significant impact on reconstructed results, making it difficult to isolate the influence of other factors on privacy leakage. 
Therefore, we use conventional gradient inversion attack methods~\cite{geipingInvertingGradientsHow2020} that do not involve the use of GANs. 

There have been several evaluations and reviews of these privacy attacks against deep learning models~\cite{heQuantifyingMitigatingPrivacy2021,liuWhenMachineLearning2021,songSystematicEvaluationPrivacy2021,liuMLDoctorHolisticRisk2022,huMembershipInferenceAttacks2022,zhangSurveyGradientInversion2022,zhangVisualPrivacyAttacks2022}.
However, we aim to evaluate the model architectures leveraging these privacy attacks.
To sum up, we utilize conventional privacy attacks~\cite{shokriMembershipInferenceAttacks2017,salemMLLeaksModelData2019,carliniMembershipInferenceAttacks2022,melisExploitingUnintendedFeature2019,songOverlearningRevealsSensitive2020,geipingInvertingGradientsHow2020} as the baseline attacks in our analysis,
for these attack methods have inspired many follow-up research works,
and they are suitable for evaluation on various models and datasets.

\section{Methodology of Evaluating the Impact of the Model Architecture on Privacy}
\label{sec:evaluate_the_impact}

In this section, we present our approach to assessing the impact of model architectures on privacy leakage.
In order to organize our study in a thorough and logical manner, We aim to answer the following research questions sequentially:

\begin{itemize}
    \itemsep-0.28em 
    \item \textbf{RQ1}: How to analyze the privacy leakage in model architectures?
    \item \textbf{RQ2}: What model architectures of CNNs and Transformers should we choose to evaluate these attacks?
    \item \textbf{RQ3}: What performance aspects should we focus on when evaluating the privacy attacks on model architectures?
    \item \textbf{RQ4}: How should we investigate what designs in model architectures contribute to privacy leakage?
\end{itemize}

In this work, we focus on classifier or feature representation models such as CNNs and Transformers, which are subject to the investigated privacy attacks. A new line of generative AI models, such as generative adversarial networks (GANs) and diffusion models, are vulnerable to different privacy attacks and thus out of the scope of this paper.
We believe our evaluation methodology could shed light on the effect of model privacy from the perspective of model architectures.

\subsection{Privacy Threat Models}
\label{sec:privacy_threat_models}

To answer the first research question (\textbf{RQ1}), we choose three prominent privacy attack methods: membership inference attacks, attribute inference attacks, and gradient inversion attacks.

\subsubsection{Membership Inference Attacks}

\textbf{Network-Based Attacks.}
Initiating a network-based membership inference attack~\cite{shokriMembershipInferenceAttacks2017,salemMLLeaksModelData2019} requires three models: the victim model $\mathcal{V}$ (the target), the shadow model $\mathcal{S}$ (the model to mimic the behavior of the victim model), and the attack model $\mathcal{A}$ (the classifier to give results whether the sample belongs to the member or non-member data).
The following paragraphs provide explanations of how the attacks work.

The first step is the attack preparation.
Since the adversary has only black-box access to the victim model $\mathcal{V}$, they can only query the model and record prediction results. 
To launch a membership inference attack, the adversary needs to create a shadow model $\mathcal{S}$, which behaves similarly to the victim model $\mathcal{V}$. 
This involves collecting a shadow dataset $\mathcal{D}_S$, usually from the same data distribution as the victim dataset $\mathcal{D}_V$. 
The shadow dataset $\mathcal{D}_S$ is then divided into two subsets: $\mathcal{D}_S^{train}$ for training and $\mathcal{D}_S^{test}$ for testing.

Once the preparation is complete, the adversary trains the attack model.
The shadow model $\mathcal{S}$ and shadow dataset $\mathcal{D}_S$ are used to train the attack model $\mathcal{A}$.
Each prediction result of a data sample from the shadow dataset $\mathcal{D}_S$ is a vector of confidence scores for each class, which is concatenated with a binary label indicating whether the prediction is correct or not.
The resulting vector, denoted as $\mathcal{P}_S^i$, is collected for all $n$ samples, forming the input set $\mathcal{P}_S = \{\mathcal{P}_S^i, i = 1, ..., n\}$ for the attack model $\mathcal{A}$.
Since $\mathcal{A}$ is a binary classifier, a three-layer MLP (multi-layer perceptron) model is employed to train it.

At last, the adversary launches the attack model inference.
The adversary queries the victim model $\mathcal{V}$ with the victim dataset $\mathcal{D}_V$ and records the prediction results, which are used as the input for the attack model $\mathcal{A}$. The attack model then predicts whether a data sample is a member or non-member data sample.

\textbf{Likelihood-Based Attacks.}
The Likelihood Ratio Attack (LiRA)~\cite{carliniMembershipInferenceAttacks2022} is a state-of-the-art attack method that employs both model posteriors and their likelihoods based on shadow models. 
In contrast to attacks relying on a single shadow model, LiRA requires the adversary to train multiple shadow models $\mathcal{S} = \{S_1, ..., S_n\}$. 
This ensures that a target sample (from the victim dataset $\mathcal{D}_V$) is included in half of the models $\mathcal{S}$ and excluded from the other half. 
The adversary then queries the shadow models with the target sample and calculates the logits for each model. 
Using these logits, the adversary calculates the probability density function to determine the likelihood ratio of the target sample, which corresponds to its membership status.

There are other kinds of membership inference attacks, including metric-based attacks and label-only attacks~\cite{songSystematicEvaluationPrivacy2021,choquette-chooLabelOnlyMembershipInference2021,liMembershipLeakageLabelOnly2021}.
Instead of using a neural network to be the attack model, metric-based attacks~\cite{songSystematicEvaluationPrivacy2021} launch the attacks using a certain metric and threshold to separate member data from non-member data.
Label-only attacks~\cite{choquette-chooLabelOnlyMembershipInference2021,liMembershipLeakageLabelOnly2021} relax the assumptions of the threat model leveraging only prediction labels as the input of the attack model.
Our study focuses on two types of membership inference attacks: network-based and likelihood-based attacks. 
We chose these two types of attacks because the network-based attack is commonly used as a baseline in many research papers, making it a conventional attack to consider. 
Additionally, the likelihood-based attack is a more recent state-of-the-art attack that has demonstrated high effectiveness, making it an important attack to evaluate as well. 
By considering these two types of attacks, we can effectively represent the performance of membership inference attacks against various victim models and gain insights into potential privacy risks associated with different machine learning models.

\subsubsection{Attribute Inference Attacks}

The goal of attribute inference attacks~\cite{melisExploitingUnintendedFeature2019,songOverlearningRevealsSensitive2020} is to extract sensitive attributes from a victim model, which may inadvertently reveal information about the training data. 
For instance, suppose the victim model is trained to classify whether a person has a beard or not. 
In that case, an adversary may infer the person's race based on the model's learned representation.

At the attack preparation stage, the victim model $\mathcal{V}$ is trained by the victim dataset $\mathcal{D}_V$ with two subsets $\mathcal{D}_V^{train}$ and $\mathcal{D}_V^{test}$ for the training and testing.

The second step is also the attack model training.
To train the attack model $\mathcal{A}$, the adversary uses an auxiliary dataset $\mathcal{D}_A^{train}$, which includes pairs of the representation $h$ and the attribute $a$, i.e., $(h, a) \in \mathcal{D}_A$.

At last, the adversary launches the attack.
The adversary takes a data sample's representation $h$ as the input and uses the attack model $\mathcal{A}$ to infer the attribute result.

\subsubsection{Gradient Inversion Attacks}

Launching the gradient inversion attack~\cite{zhuDeepLeakageGradients2019,zhaoIDLGImprovedDeep2020,geipingInvertingGradientsHow2020} involves solving an optimization problem, which aims to minimize the difference between the calculated model gradients and the original model gradients. The optimization process continues for a certain number of iterations, after which the input data sample can be reconstructed.

The adversary operates within a federated learning scenario.
In the attack preparation stage, the adversary operates from the central server, aggregating model gradients to create a centralized model. 
Since the adversary has access to the communication channels used during the federated learning process, they can retrieve the model gradients and prepare to extract sensitive information from the training samples. 
This allows the adversary to launch attacks against the federated learning system.

In the step of gradient reconstruction, the aggregated model gradients are denoted as $\nabla _\theta \mathcal{L}_\theta (x, y)$, where $\theta$ is the model parameters, $x$ and $y$ are the original input image and its ground truth in a local client, and $\mathcal{L}$ represents the cost function for the model. 
To initiate the reconstruction process, the adversary generates a dummy image $x^*$.
The adversary tries to minimize this cost function: $\arg \min_x || \nabla _\theta \mathcal{L}_\theta (x, y) - \nabla _\theta \mathcal{L}_\theta (x^*, y) ||^2$.
The dummy image $x^*$ is reconstructed to resemble $x$ closely.

\subsection{CNNs vs Transformers}
\label{sec:cnn_vs_transformers}

To answer the second research question (\textbf{RQ2}), we investigate the privacy of two mainstream architectures: CNNs and Transformers.
We carefully select several popular CNNs and Transformers for the attacks to analyze the privacy leakage.

For CNNs, we choose ResNets~\cite{heDeepResidualLearning2016} as baseline models, which are known for incorporating residual blocks and have become widely used in various computer vision tasks. 
We specifically select ResNet-50 (23.52 million parameters) and ResNet-101 (42.51 million parameters) to represent CNN architectures in our analysis.
Regarding Transformers, we focus on Swin Transformers~\cite{liuSwinTransformerHierarchical2021}, which have gained attention for their innovative design incorporating attention modules and shifted window mechanisms.
We analyze Swin-T (27.51 million parameters) and Swin-S (48.80 million parameters) as representatives of Transformer architectures.

To ensure fair comparisons, we organize the evaluation of the four models based on their parameter sizes, grouping models with similar parameter sizes together. 
This approach allows us to compare models that exhibit comparable task performances while considering their architectural differences. 
Specifically, we compare ResNet-50 with Swin-T, as they have similar parameter sizes, and we compare ResNet-101 with Swin-S for the same reason.

Apart from model architectures, 
training Transformers requires a modernized training procedure compared to training traditional CNNs~\cite{vaswaniAttentionAllYou2017,liuSwinTransformerHierarchical2021}.
To ensure fair comparisons, we employ the same training recipe for both CNNs and Transformers in each comparison.

\begin{table}[t]
  \scriptsize
  \caption{Training recipes for privacy attacks.}
  \label{tab:recipe}
  \centering
  \resizebox{\linewidth}{!}{
  \begin{tabular}{lc|lc}
    \toprule
    Config    & Param & Config    & Param\\
    \midrule
    \multicolumn{4}{l}{for all models of membership inference} \\
    \midrule
    optimizer & AdamW  &  training epochs & 300 \\
    learning rate & 0.001 & batch size & 256\\
    weight decay & 0.05 & mixup & 0.8\\
    optimizer momentum & $\beta_1, \beta_2$ = 0.9, 0.999 & cutmix & 1.0\\
    learning rate schedule & cosine annealing & label smoothing & None\\
    \midrule
    \multicolumn{4}{l}{for all models of attribute inference} \\
    \midrule
    optimizer & SGD    & batch size & 256 \\
    learning rate & 0.01 & training epochs & 100\\
    weight decay & 0.0005 & random augmentation & None\\
    optimizer momentum & 0.9 & & \\
    \midrule
    \multicolumn{4}{l}{for gradient inversion} \\
    \midrule
    cost function & similarity & total iteration & 3000\\
    optimizer & Adam & total variance & 0.0001\\
    learning rate & 0.1 & & \\
    \bottomrule
  \end{tabular}}
\end{table}

\section{Evaluation on Privacy Attacks with CNNs and Transformers}
\label{sec:experiments}

This section addresses the third research question (\textbf{RQ3}) and comprehensively analyzes the experimental settings and results. 
We aim to compare the attack performance under various metrics to gain a deeper understanding of the findings.
We also focus on the performance differences and overfitting differences between CNNs and Transformers.

\subsection{Settings}
\label{sec:settings}

\textbf{Datasets.} Our experiments evaluate the privacy leakage under these four datasets.

\begin{itemize}
    \itemsep-0.28em 
    \item \textbf{CIFAR10~\cite{krizhevskyLearningMultipleLayers2009}} consists of 60,000 color images with dimensions of $32 \times 32$ pixels. 
    It is organized into ten classes, with each class containing 6,000 images. 
    The dataset covers a wide range of general object categories.
    \item \textbf{CIFAR100~\cite{krizhevskyLearningMultipleLayers2009}} is similar to CIFAR10 but with 100 classes.
    Each class has 600 images and represents a specific general object category.
    \item \textbf{ImageNet1K~\cite{dengImageNetLargescaleHierarchical2009}} is a widely used dataset in computer vision containing over 1 million labeled images, covering 1,000 different classes. 
    The dataset encompasses a diverse range of objects and scenes.
    \item \textbf{CelebA~\cite{liuDeepLearningFace2015}} contains over 200,000 face images, each with 40 binary attributes.
\end{itemize}

In membership inference attacks, we use two datasets, CIFAR10 and CIFAR100, for their popularity in prior attack research~\cite{salemMLLeaksModelData2019,carliniMembershipInferenceAttacks2022,liuMLDoctorHolisticRisk2022}.
For network-based attacks, each dataset is evenly split into four subsets for training and testing both the victim and shadow models to ensure a fair evaluation~\cite{salemMLLeaksModelData2019}.
For likelihood-based attacks, we follow the training settings in~\cite{carliniMembershipInferenceAttacks2022}.
This involves dividing the training data into multiple subsets, ensuring that a specific training sample is present in only half of these subsets.

For attribute inference attacks, we utilize the CelebA dataset, which provides rich attribute labeling. 
This allows us to identify hidden attributes accurately. 
In our experiments, we focus on inferring the race attribute while using the gender attribute as the classification goal for the victim model. 
We randomly select 20,000 images from CelebA and evenly split them into four subsets for training and testing both the victim model and the attribute inference attack model.

In gradient inversion attacks, we employ the CIFAR10 and ImageNet1K datasets for low/high-resolution reconstruction.
These attacks are conducted in a federated learning scenario to reconstruct the training batch. 
We randomly select a subset of images from these datasets to evaluate the attacks, ensuring a representative assessment of the attack performance.

In a nutshell, all these datasets are benchmark datasets for evaluating privacy attacks~\cite{heQuantifyingMitigatingPrivacy2021,carliniMembershipInferenceAttacks2022,liuMLDoctorHolisticRisk2022,geipingInvertingGradientsHow2020,yinSeeGradientsImage2021}.
These datasets cover a wide range of objects with different numbers of classes, making them sufficient for our evaluation.

\textbf{Victim Models.}
To ensure fair comparisons, we evaluate the performances of CNNs and Transformers with similar parameter sizes.
For all attacks, we select two groups of models based on their parameter sizes. 
These groups include ResNet-50 and ResNet-101 as CNN-based models and Swin-T and Swin-S as Transformer-based models.
We train these models to reach more than 0.99 training accuracy and output testing accuracy (i.e., task accuracy) results in the experiments.

\textbf{Attack Models.} 
For membership inference attacks, we employ a three-layer MLP model to infer membership information. 
Attribute inference attacks utilize a two-layer MLP model to uncover learned representations and potentially infer sensitive information. 
In gradient inversion attacks, we optimize input and generate gradients to reconstruct the original data, revealing private information.

\textbf{Evaluation Metrics.} 
We evaluate the performance of different attacks using specific metrics. 
For membership inference attacks, we consider attack accuracy ($\uparrow$), ROC curve ($\uparrow$), AUC ($\uparrow$), and TPR at low FPR ($\uparrow$).
For attribute inference attacks, we assess effectiveness using attack accuracy ($\uparrow$) and macro-F1 score ($\uparrow$). 
Regarding gradient inversion attacks, we use multiple metrics to evaluate the quality of reconstruction results. 
These metrics include mean square error (MSE $\downarrow$), peak signal-to-noise ratio (PSNR $\uparrow$), learned perceptual image patch similarity (LPIPS $\downarrow$)~\cite{zhangUnreasonableEffectivenessDeep2018}, and structural similarity (SSIM $\uparrow$)~\cite{wangImageQualityAssessment2004}. 
Note that "$\uparrow$" means the higher metric corresponds to the higher attack performance, while $\downarrow$ means the lower metric leads to the higher attack performance.

\textbf{Training Settings for Privacy Attacks.} \Cref{tab:recipe} illustrates the training configurations for membership inference attacks, attribute inference attacks and gradient inversion attacks.

\begin{table}[t]
  \scriptsize
  \centering
  \caption{Results for network-based membership inference attacks.}
    \resizebox{\linewidth}{!}{
    \begin{tabular}{c|cc|cc}
    \toprule
    & \multicolumn{2}{c|}{CIFAR10} & \multicolumn{2}{c}{CIFAR100} \\
    \cmidrule{2-5}         
    & \multicolumn{1}{c}{Task acc $\uparrow$} & \multicolumn{1}{c|}{Attack acc $\uparrow$} & \multicolumn{1}{c}{Task acc $\uparrow$} & \multicolumn{1}{c}{Attack acc $\uparrow$} \\
    \midrule
    ResNet-50  & 0.8220 $\pm$ 0.0023 & 0.6385 $\pm$ 0.0078 & 0.5288 $\pm$ 0.0083 & 0.8735 $\pm$ 0.0029 \\
    Swin-T     & 0.8335 $\pm$ 0.0042 & 0.6904 $\pm$ 0.0052 & 0.5632 $\pm$ 0.0056 & 0.9340 $\pm$ 0.0030 \\
    \midrule
    ResNet-101 & 0.8301 $\pm$ 0.0037 & 0.6317 $\pm$ 0.0063 & 0.5313 $\pm$ 0.0074 & 0.8607 $\pm$ 0.0034 \\
    Swin-S     & 0.8258 $\pm$ 0.0039 & 0.6405 $\pm$ 0.0075 & 0.5665 $\pm$ 0.0059 & 0.9357 $\pm$ 0.0039 \\
    \bottomrule
    \end{tabular}}%
  \label{tab:network_mia_results}%
\end{table}%

\subsection{Evaluation on Membership Inference Attacks}
\label{sec:exp_mia}

The results presented in Table \ref{tab:network_mia_results} offer the performance of network-based membership inference attacks on CIFAR10 and CIFAR100 datasets. 
The task accuracy scores of the victim models on CIFAR10, which are approximately 0.82 for both CNNs and Transformers, indicate that these models exhibit competitive task performance with similar overfitting levels.
Notably, the attack accuracy on CIFAR10 reveals that Transformers exhibit more privacy leakage within each group than CNN models. 
Similar findings are observed on CIFAR100, suggesting that Transformers consistently exhibit higher vulnerability to membership inference attacks compared to CNN models.

It is worth mentioning that the dataset used in this study is divided into four equally sized subsets for training and testing the victim and attack models. 
Consequently, the task accuracy of the victim models on CIFAR10 and CIFAR100 might be lower than expected in a standard CIFAR10 or CIFAR100 classification task. 
This does not affect the performance of the attacks, and this phenomenon has been acknowledged in prior research \cite{heQuantifyingMitigatingPrivacy2021,kayaWhenDoesData2021,shejwalkarMembershipPrivacyMachine2021,carliniMembershipInferenceAttacks2022}.
This also applies to later experimental results.

\begin{figure}[t]
  \scriptsize
  \begin{subfigure}[htbp]{.23\textwidth}
    \centering
    \includegraphics[width=\linewidth]{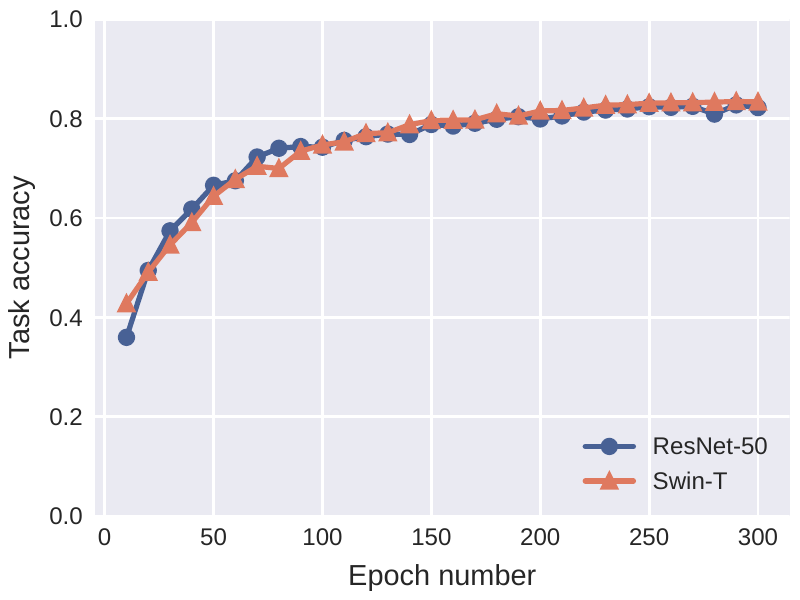}
    \caption{Performance of victim models on CIFAR10}
    \label{fig:c10_victim_acc_multi}
  \end{subfigure}
  \hfill
  \begin{subfigure}[htbp]{.23\textwidth}
    \centering
    \includegraphics[width=\linewidth]{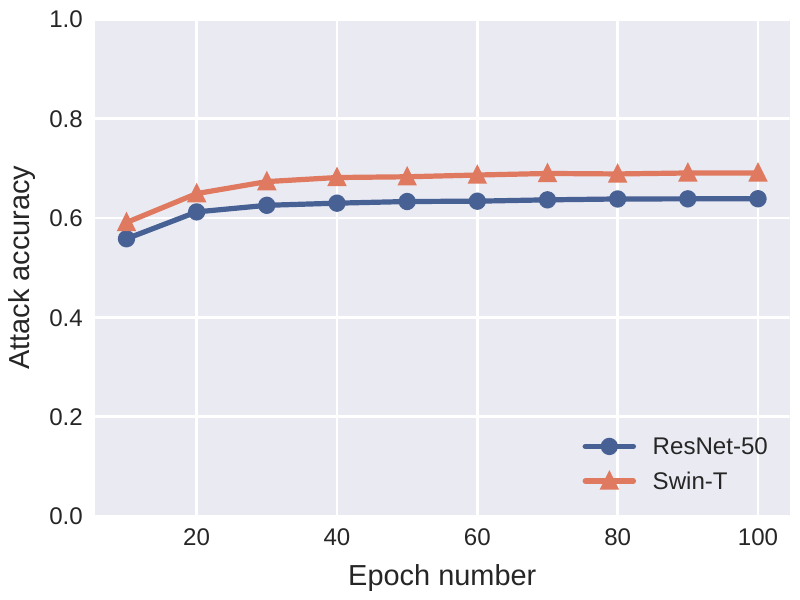}
    \caption{Performance of privacy attack on CIFAR10}
    \label{fig:c10_attack_acc_multi}
  \end{subfigure}
  \caption{The performance of membership inference attacks against ResNet-50 and Swin-T on CIFAR10 under different numbers of epochs.}
  \label{fig:mia_acc_multi}
\end{figure}

\Cref{fig:mia_acc_multi} illustrates the task accuracy and attack accuracy of ResNet-50 and Swin-T across multiple epochs on CIFAR10.
The task accuracy of ResNet-50 starts relatively low but exhibits a rapid increase over time. Eventually, both ResNet-50 and Swin-T can achieve task accuracy scores of approximately 0.82.
Regarding attack accuracy, the plot demonstrates that the attack on Swin-T consistently outperforms the attack on ResNet-50.

\begin{table}[t]
  \scriptsize
  \centering
  \caption{Results of likelihood-based membership inference attacks.}
    \resizebox{\linewidth}{!}{
    \begin{tabular}{c|c|c|ccc}
    \toprule
    \multicolumn{1}{r}{} &       & Task acc $\uparrow$ & AUC $\uparrow$  & TPR@0.1\%FPR $\uparrow$ & Attack acc  $\uparrow$ \\
    \midrule
    \multirow{4}[4]{*}{\begin{turn}{-90}CIFAR10\end{turn}} & ResNet-50 & 0.8716 $\pm$ 0.0035 & 0.6446 $\pm$ 0.0276 & 3.15\% $\pm$ 0.25\% & 0.6009 $\pm$ 0.0163 \\
          & Swin-T & 0.8630 $\pm$ 0.0017 & 0.7384 $\pm$ 0.0029 & 3.68\% $\pm$ 0.33\% & 0.6553 $\pm$0.0027 \\
\cmidrule{2-6}          & ResNet-101 & 0.8708 $\pm$ 0.0043 & 0.6671 $\pm$ 0.0107 & 3.33\% $\pm$ 0.41\% & 0.6090 $\pm$ 0.0079 \\
          & Swin-S & 0.8636 $\pm$ 0.0036 & 0.7392 $\pm$ 0.0054 & 3.75\% $\pm$ 0.37\% & 0.6576 $\pm$ 0.0045 \\
    \midrule
    \multirow{4}[4]{*}{\begin{turn}{-90}CIFAR100\end{turn}} & ResNet-50 & 0.5632 $\pm$ 0.0032 & 0.9431 $\pm$ 0.0005 & 23.75\% $\pm$ 2.09\% & 0.8524 $\pm$ 0.0009 \\
          & Swin-T & 0.6001 $\pm$ 0.0033 & 0.9756 $\pm$ 0.0003 & 28.52\% $\pm$ 1.54\% & 0.9112 $\pm$ 0.0010 \\
\cmidrule{2-6}          & ResNet-101 & 0.5654 $\pm$ 0.0042 & 0.9379 $\pm$ 0.0021 & 21.25\% $\pm$ 1.55\% & 0.8484 $\pm$ 0.0036 \\
          & Swin-S & 0.5900 $\pm$ 0.0029 & 0.9639 $\pm$ 0.0006 & 31.02\% $\pm$ 1.74\% & 0.8978 $\pm$ 0.0014 \\
    \bottomrule
    \end{tabular}}%
  \label{tab:likelihood_mia_results}%
\end{table}%

\begin{figure}[t]
  \scriptsize
  \begin{subfigure}[htbp]{.23\textwidth}
    \centering
    \includegraphics[width=\linewidth]{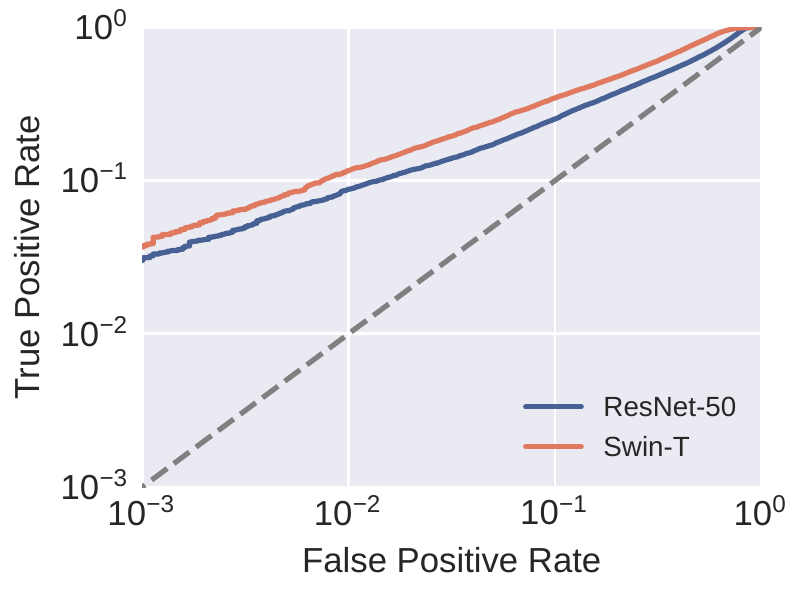}
    \caption{ResNet-50 vs Swin-T on CIFAR10}
    \label{fig:mia_roc_cifar10_r50st}
  \end{subfigure}
  \hfill
  \begin{subfigure}[htbp]{.23\textwidth}
    \centering
    \includegraphics[width=\linewidth]{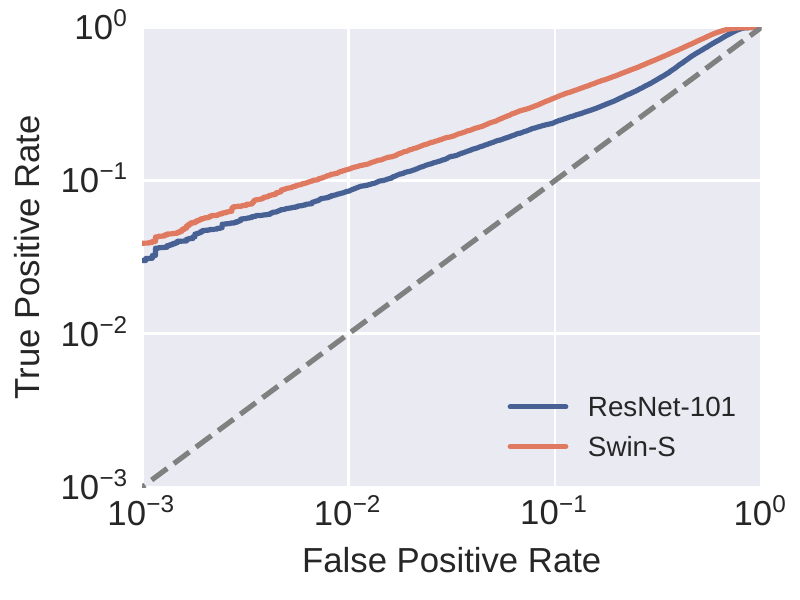}
    \caption{ResNet-101 vs Swin-S on CIFAR10}
    \label{fig:mia_roc_cifar10_r101ss}
  \end{subfigure}
  \hfill
  \medskip
  \begin{subfigure}[htbp]{.23\textwidth}
    \centering
    \includegraphics[width=\linewidth]{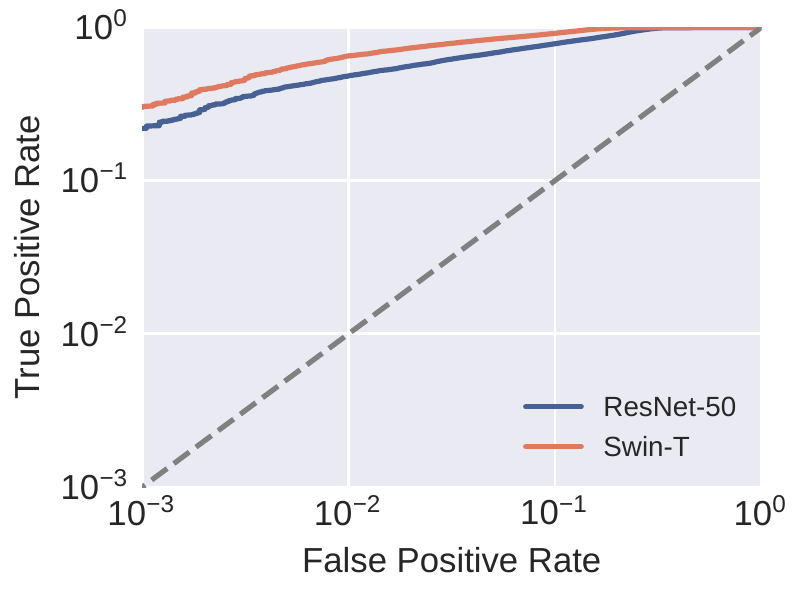}
    \caption{ResNet-50 vs Swin-T on CIFAR100}
    \label{fig:mia_roc_cifar100_r50st}
  \end{subfigure}
  \hfill
  \begin{subfigure}[htbp]{.23\textwidth}
    \centering
    \includegraphics[width=\linewidth]{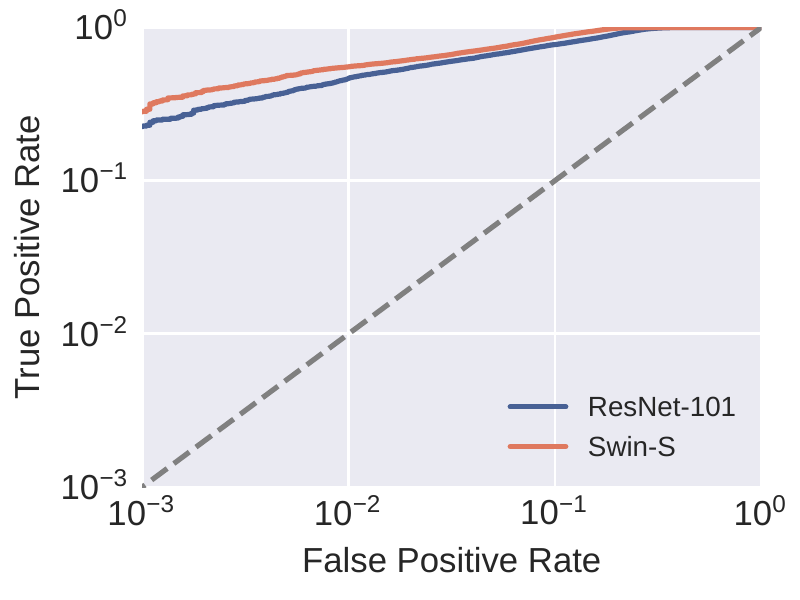}
    \caption{ResNet-101 vs Swin-S on CIFAR100}
    \label{fig:mia_roc_cifar100_r101ss}
  \end{subfigure}
  \caption{ROC curves for membership inference attacks on CIFAR10 and CIFAR100. Comparisons between CNNs and Transformers.}
  \label{fig:mia_roc}
\end{figure}

\Cref{tab:likelihood_mia_results} displays the outcomes of likelihood-based membership inference attacks on CIFAR10 and CIFAR100. 
The table includes task accuracy and attack performance, the same as the evaluation for network-based attacks.
To assess attack performance more comprehensively, we adopt the methodology proposed in \cite{carliniMembershipInferenceAttacks2022} and incorporate additional metrics such as AUC, TPR@0.1\%FPR, besides attack accuracy.
The victim models achieve task accuracy scores of approximately 0.86 on CIFAR10 and 0.56 on CIFAR100.
Consistently, the attack performance metrics highlight that Transformers are more vulnerable to membership inference attacks compared to CNNs in terms of any attack metric.
We also present the attack performance through ROC curves in~\Cref{fig:mia_roc}. 
These curves demonstrate that Transformers yield higher ROC curves, signifying their better attack performance.

The privacy leakage varies on different models and datasets.
Similar to~\cite{shokriMembershipInferenceAttacks2017,salemMLLeaksModelData2019}, 
we analyze the overfitting levels of victims models in~\Cref{fig:mia_overfitting_level}.
The overfitting level indicates the accuracy difference of a model between its training and inference.
\Cref{fig:mia_overfitting_level} illustrates the results of Transformers and CNNs in CIFAR10 and CIFAR100.
We conclude that a more overfitted model comes with higher membership inference attack accuracy.
More importantly, at the same overfitting level, Transformers always suffer from higher attack accuracy.

In conclusion, the results strongly suggest that Transformers are more susceptible to network-based and likelihood-based membership inference attacks compared to CNNs. 
Transformers consistently demonstrate higher vulnerability to membership inference attacks across various metrics.

\begin{table}[t]
\scriptsize
\centering
  \caption{Results of attribute inference attacks on CelebA.}
    \begin{tabular}{c|c|cc}
    \toprule
    & Task acc $\uparrow$ & Attack acc $\uparrow$ & Macro F1 $\uparrow$ \\
    \midrule
    ResNet-50 & 0.6666 $\pm$ 0.0020 & 0.6854 $\pm$ 0.0015 & 0.3753 $\pm$ 0.0012 \\
    Swin-T & 0.6587 $\pm$ 0.0023 & 0.7312 $\pm$ 0.0014 & 0.5530 $\pm$ 0.0019 \\
    \midrule
    ResNet-101 & 0.6431 $\pm$ 0.0029 & 0.6291 $\pm$ 0.0023 & 0.4262 $\pm$ 0.0009 \\
    Swin-S & 0.6569 $\pm$ 0.0024 & 0.7369 $\pm$ 0.0036 & 0.5536 $\pm$ 0.0015 \\
    \bottomrule
    \end{tabular}%
  \label{tab:aia_acc}%
\end{table}%

\begin{figure}[t]
  \scriptsize
  \begin{subfigure}[htbp]{.23\textwidth}
    \centering
    \includegraphics[width=\linewidth]{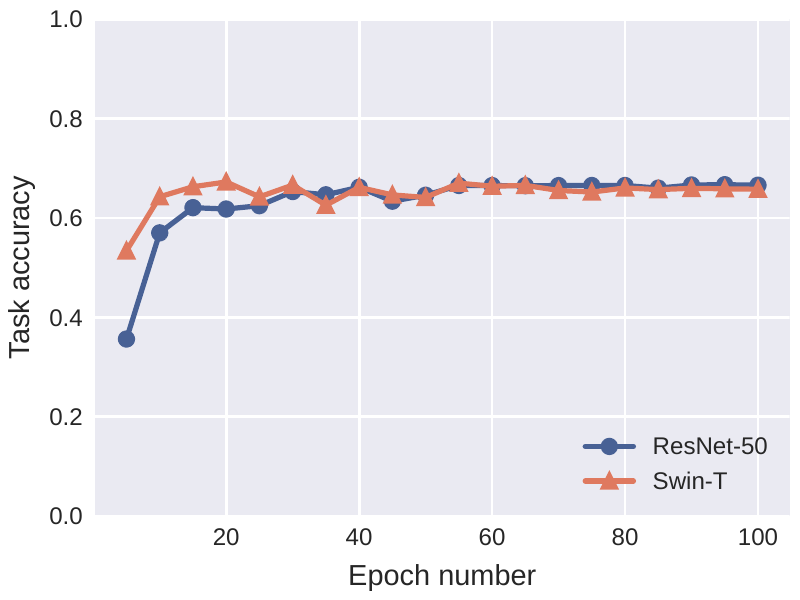}
    \caption{Performance of victim models on CelebA}
    \label{fig:celeba_aia_task_acc_multi}
  \end{subfigure}
  \hfill
  \begin{subfigure}[htbp]{.23\textwidth}
    \centering
    \includegraphics[width=\linewidth]{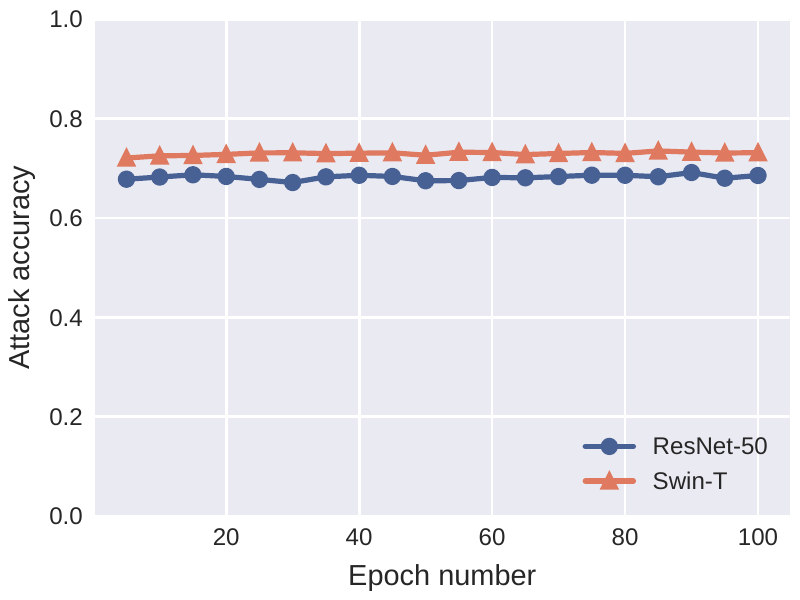}
    \caption{Performance of privacy attack on CelebA}
    \label{fig:celeba_aia_attack_acc_multi}
  \end{subfigure}
  \caption{The performance of attribute inference attacks against ResNet-50 and Swin-T on CelebA under different numbers of epochs.}
  \label{fig:aia_acc_multi}
\end{figure}

\subsection{Evaluation on Attribute Inference Attacks}
\label{sec:exp_aia}

\Cref{tab:aia_acc} presents the results of attribute inference attacks on CelebA.
Similarly to membership inference attacks, we categorize CNN and Transformer models into two groups based on their parameter sizes.
The table shows that within each group, CNNs and Transformers achieve similar task accuracy scores, approximately 0.65. 
However, when considering the attack accuracy and Macro F1 score, Transformers consistently outperform CNNs.
The results from attribute inference attacks align with our previous findings from membership inference attacks, emphasizing the increased vulnerability of Transformer models to privacy attacks. 

\Cref{fig:aia_acc_multi} presents the performance of task accuracy and attack accuracy in attribute inference attacks using ResNet-50 and Swin-T on CelebA over 100 epochs.
\Cref{fig:celeba_aia_task_acc_multi} shows that although ResNet-50 and Swin-T models start with different task accuracy scores, both models gradually converge to similar task accuracy scores of around 0.65. 
However, when we examine the attack accuracy in Figure \ref{fig:celeba_aia_attack_acc_multi}, the attack accuracy on Swin-T consistently outperforms that on ResNet-50 throughout the 100 epochs. 
This reveals that Transformers like Swin-T are more vulnerable to attribute inference attacks than ResNet-50 from the start of the attack training to the end.

\begin{figure}[t]
  \scriptsize
  \begin{subfigure}[htbp]{.23\textwidth}
    \centering
    \includegraphics[width=\linewidth]{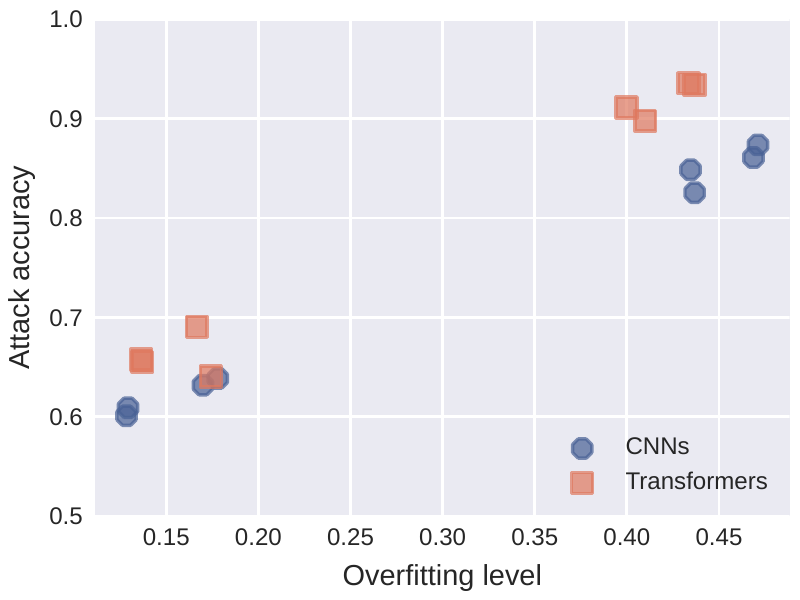}
    \caption{Membership inference}
    \label{fig:mia_overfitting_level}
  \end{subfigure}
  \hfill
  \begin{subfigure}[htbp]{.23\textwidth}
    \centering
    \includegraphics[width=\linewidth]{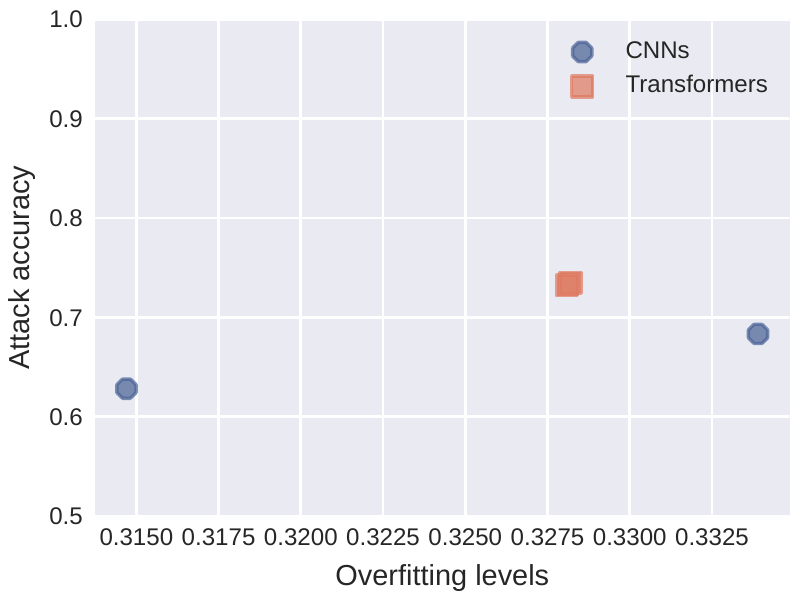}
    \caption{Attribute inference}
    \label{fig:aia_overfitting_level}
  \end{subfigure}
  \caption{The performance of privacy attacks against both CNNs and Transformers with various models and datasets under different overfitting levels.}
  \label{fig:overfitting_level}
\end{figure}

We further analyze the relationship between the attack performance and the overfitting levels of victim models in~\Cref{fig:aia_overfitting_level}.
We have made a similar discovery to the previous evaluation: Transformers suffer from higher attack accuracy than CNNs when the victim models are at the same overfitting level. 

\begin{table}[t]
  \scriptsize
  \centering
  \caption{The results of gradient inversion attacks on CNNs and Transformers on CIFAR10.}
    \resizebox{\linewidth}{!}{
    \begin{tabular}{c|cccc}
    \toprule
    & MSE $\downarrow$ & PSNR $\uparrow$ & LPIPS $\downarrow$ & SSIM $\uparrow$ \\
    \midrule
    ResNet-50 & 1.3308 $\pm$ 0.6507 & 11.30 $\pm$ 2.24 & 0.1143 $\pm$ 0.0403 & 0.0946 $\pm$ 0.0989 \\
    Swin-T & 0.0069 $\pm$ 0.0071 & 36.24 $\pm$ 5.21 & 0.0012 $\pm$ 0.0016 & 0.9892 $\pm$ 0.0118 \\
    \midrule
    ResNet-101 & 1.2557 $\pm$ 0.6829 & 11.58 $\pm$ 2.16 & 0.1461 $\pm$ 0.1012 & 0.0784 $\pm$ 0.0675 \\
    Swin-S & 0.0063 $\pm$ 0.0083 & 37.85 $\pm$ 6.15 & 0.0016 $\pm$ 0.0028 & 0.9878 $\pm$ 0.0128 \\
    \bottomrule
    \end{tabular}}%
  \label{tab:inv_comp_results}%
\end{table}%

\begin{figure}[t]
    \scriptsize
    \begin{subfigure}[htbp]{.5\textwidth}
      \centering
      \includegraphics[width=0.7\linewidth]{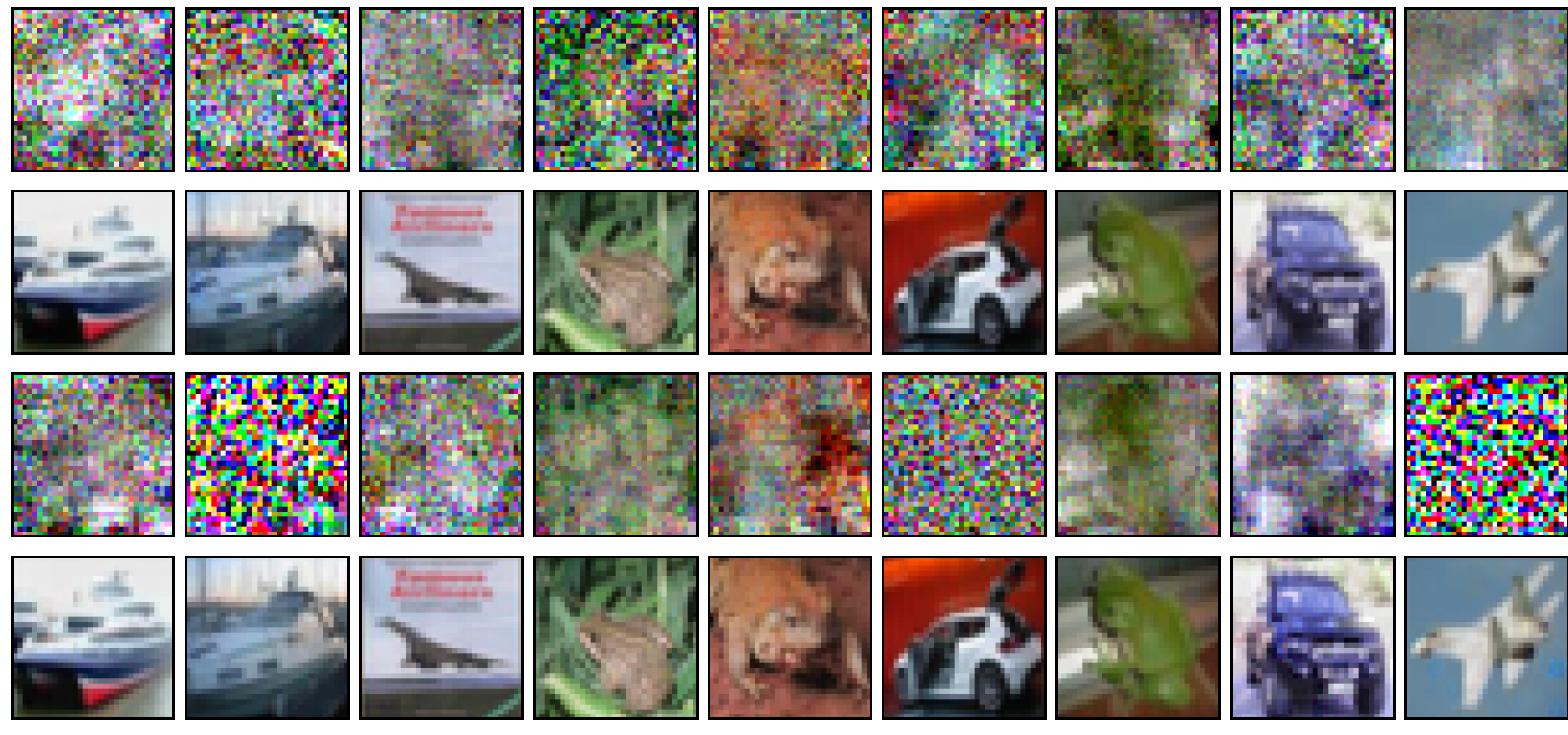}
      \caption{CIFAR10 with 3000 iterations.}
    \label{fig:inv_cnn_trans_comp}
    \end{subfigure}
    \hfill
    \begin{subfigure}[htbp]{.5\textwidth}
      \centering
      \includegraphics[width=0.7\linewidth]{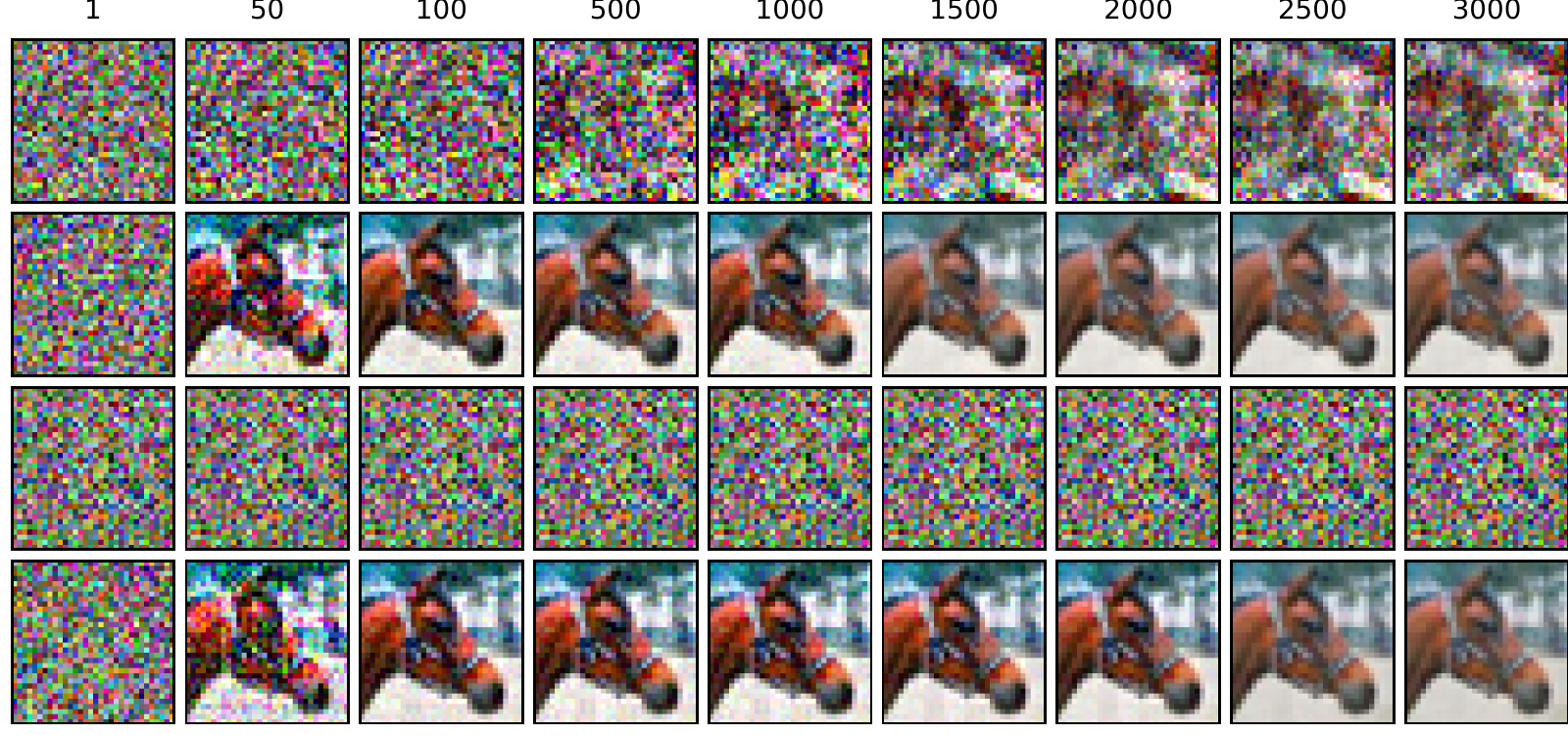}
      \caption{CIFAR10 with different iteration numbers (i.e. 1, 50, 100, 500, 1000, 1500, 2000, 2500, 3000).}
    \label{fig:inv_multi_iter}
    \end{subfigure}
    \hfill
    \begin{subfigure}[htbp]{.5\textwidth}
      \centering
      \includegraphics[width=0.7\linewidth]{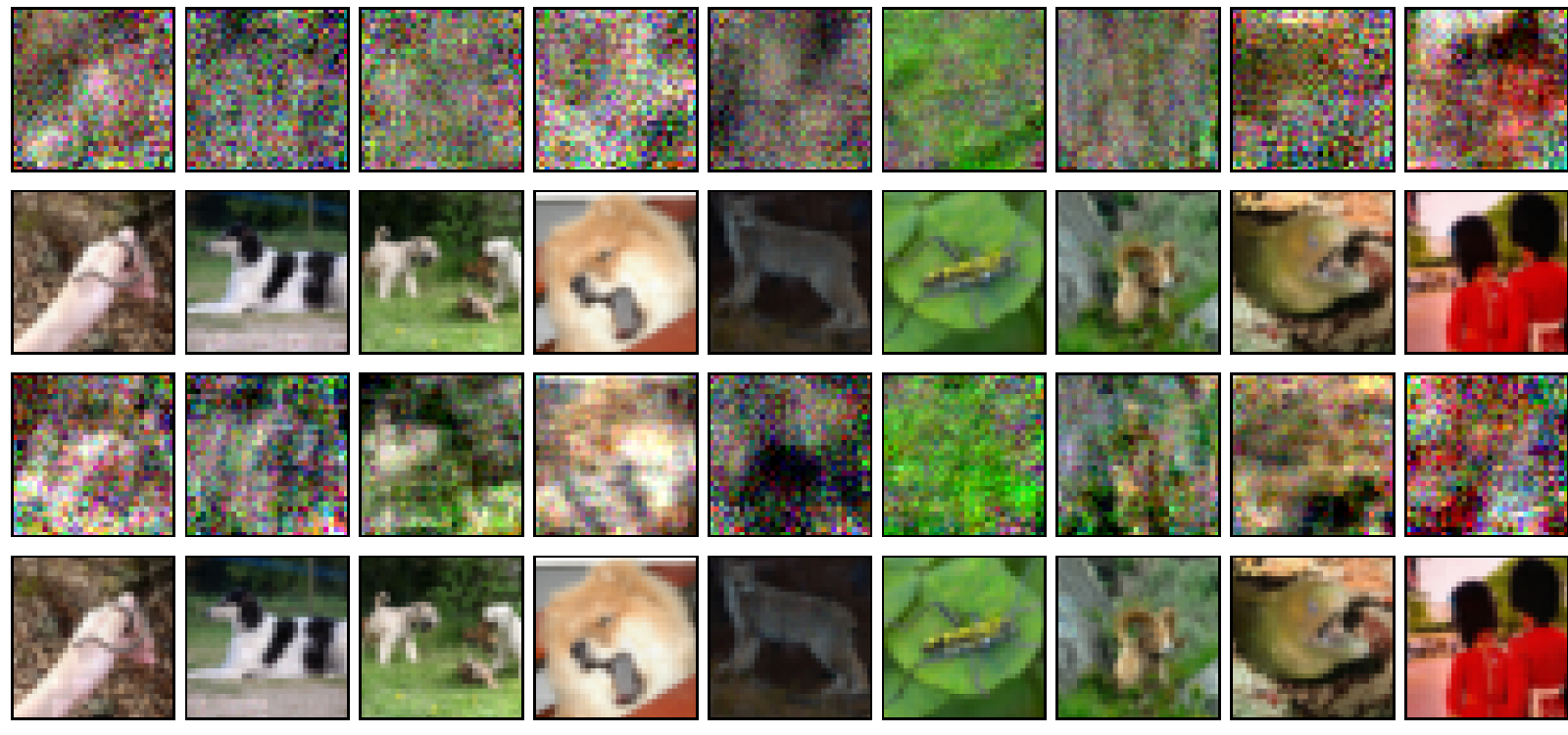}
      \caption{ImageNet1K with 3000 iterations.}
    \label{fig:imagenet}
    \end{subfigure}
    \hfill
    \caption{The performance of gradient inversion attacks on CNNs and Transformers. From the top row to the bottom in each subfigure are ResNet-50, Swin-T, ResNet-101, and Swin-S.}
    \label{fig:inv}
\end{figure}

\subsection{Evaluation on Gradient Inversion Attacks}
\label{sec:exp_gradient_inversion}

\Cref{tab:inv_comp_results} presents the results of gradient inversion attacks on CNNs and Transformers using CIFAR10. 
Similar to our previous evaluations, we still compare these two model architectures in groups.
The attacks are evaluated by multiple metrics, which measure reconstruction results between ground truth images and reconstruction images.
The table clearly shows that the attacks on Transformers outperform the attacks on CNNs by a significant margin.
\Cref{fig:inv_cnn_trans_comp} provides examples of the reconstruction results.
The attacks on CNNs fail to generate high-quality reconstruction images, whereas the attacks on Transformers produce remarkably accurate reconstructions that closely resemble the originals.

\Cref{fig:inv_multi_iter} provides the reconstruction results of the gradient inversion attacks over multiple iterations. 
It demonstrates the transformation of a raw dummy image towards a reconstruction that closely resembles the original image as the attack training continues.
The reconstruction results reveal the varying degrees of success achieved by the attacks on different models.
In the case of ResNet-50, the reconstructed image shows limited resemblance to the original image.
When attacking ResNet-101, the reconstruction result fails to capture any meaningful information.
On the other hand, when targeting Transformers such as Swin-T and Swin-S, the attacks yield highly accurate reconstruction results since early iterations.

In the evaluation of gradient inversion attacks on ImageNet1K, we still compare the performance of ResNet-50, ResNet-101, Swin-T, and Swin-S models. 
Randomly selected images from ImageNet1K are used to generate reconstruction results, shown in~\Cref{fig:imagenet}.
Similar to the previous experiments on CIFAR10, the attacks on ResNet variants have limited success in reconstructing the original images. 
In contrast, the attacks on Transformer models (Swin-T and Swin-S) yield significantly better reconstruction results. 

These findings highlight the higher vulnerability of Transformers to gradient inversion attacks compared to CNNs.
This raises questions about specific architectural features in Transformers that contribute to their increased vulnerability to privacy attacks.
We further analyze the vulnerability of architectural features in~\Cref{sec:which_arch}.

\section{Which Architectural Features Can Lead to Higher Privacy Leakage?}
\label{sec:which_arch}

In this section, we answer the fourth research question (\textbf{RQ4}).
We delve into analyzing architectural features that can potentially lead to privacy leakage. 
Through a comprehensive analysis, we first examine the impact of partitioning a transformer model on privacy vulnerabilities.
Then, we investigate the influence of various micro designs on model privacy and conduct an ablation study on these micro designs.

\begin{figure}[t]
  \scriptsize
  \centering
  \includegraphics[width=0.7\linewidth]{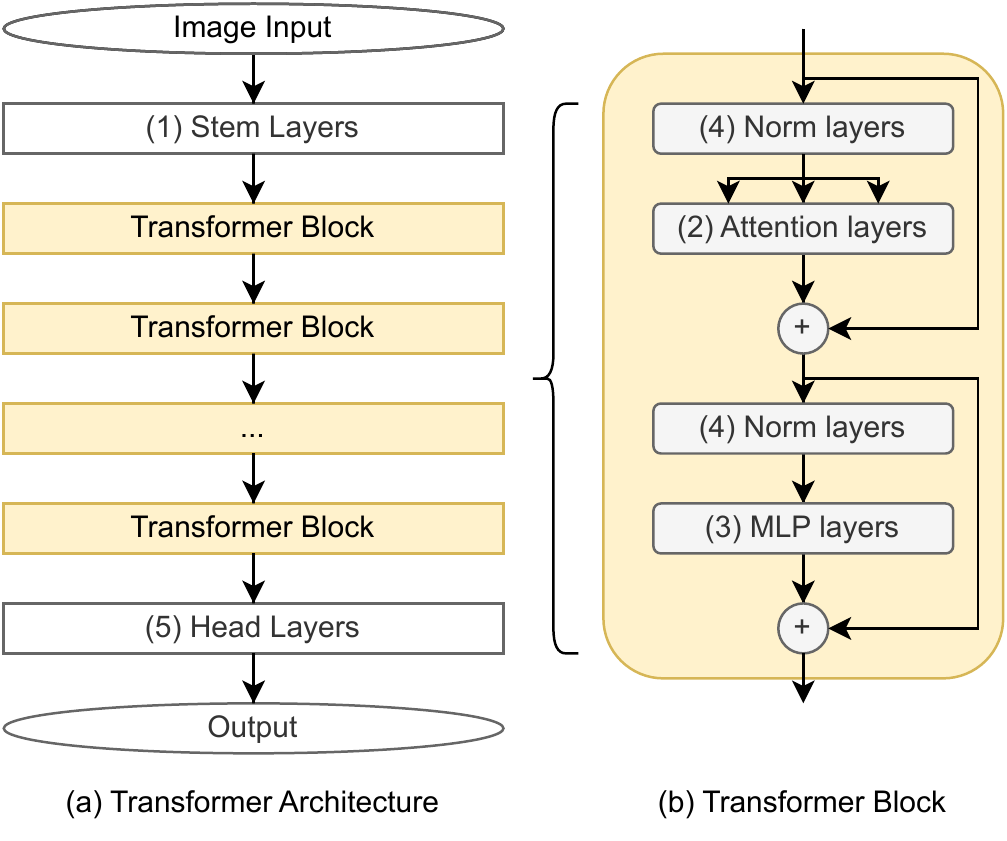}
  \caption{An illustration of a Vision Transformer architecture and the modules we focus on evaluating. Numbers (1) - (5) denote the modules we evaluate in gradient inversion attacks. (a): The whole model architecture. (b): Detailed architecture of each Transformer block.}
  \label{fig:transformer_arch}
\end{figure}

\subsection{Segmenting a Transformer Model to Analyze the Privacy Leakage}
\label{sec:segmenting_trans}

One of the key distinctions between a CNN and a Transformer lies in the design of their respective blocks. 
In this section, we focus on segmenting a Transformer-based model and assessing the potential influence of specific layers on privacy leakage. 
As there are only prediction results and model representations from membership inference attacks and attribute inference attacks, it is difficult to evaluate some specific layers using these attacks.
However, gradient inversion attacks offer a more precise means of evaluation, as they require a comprehensive list of gradients from each layer of the victim model.

In our analysis, we employ ViT-B as the victim model.
This is because ViT models are one of the first Transformers in computer vision. 
Additionally, ViT models incorporate several common modules shared across various Transformer architectures. 
We utilize gradient inversion attacks as our chosen attack method. 
Rather than supplying all the gradients to the attacks, we selectively provide specific gradients for evaluation purposes.
The architecture of a Vision Transformer (an example of ViT-B) is depicted in~\Cref{fig:transformer_arch}. 
To facilitate our assessment, we divide the model into five distinct modules based on their layer designs. Subsequently, we evaluate the influence of gradients obtained from each module individually.
If the attack using gradients from Module A yields a higher attack accuracy compared to the attack using gradients from Module B, it suggests that Module A is more likely to reveal more information about the data samples than Module B. 

\begin{itemize}
    \itemsep-0.28em 
    \item Module 1: Stem layers. This module receives the model's input and has patch embedding and position embedding layers.
    \item Module 2: Attention layers. They are in the Transformer block, and this module is the main reason why Transformers are different from CNNs.
    \item Module 3: MLP layers. They are also in the Transformer block.
    \item Module 4: Norm layers. They are located right before the attention layers and MLP layers. LayerNorm is often used as Norm layers in Transformers.
    \item Module 5: Head layers. They are the last few layers for producing the output of the model. A few fully connected layers could be used as head layers.
\end{itemize}

\begin{table}[t]
  \scriptsize
  \caption{The performance of gradient inversion attacks when segmenting ViT-B to make a selection of gradients.}
  \label{tab:gradient_sel_vitb}
  \centering
  \begin{tabular}{lcccc}
    \toprule
    Layers & Num of layers & Params & MSE $\downarrow$ & PSNR $\uparrow$ \\
    \midrule
    All       & 152 & 85.65M & 0.0007 $\pm$ 0.0003 & 43.70 $\pm$ 1.84 \\
    Stem      & 4   & 0.59M & 0.0000 $\pm$ 0.0000 & 67.43 $\pm$ 5.03 \\
    Attention & 48  & 28.34M & 0.0020 $\pm$ 0.0009 & 39.61 $\pm$ 2.76 \\
    MLP       & 48  & 56.66M & 0.0036 $\pm$ 0.0016 & 36.98 $\pm$ 2.59 \\
    Norm      & 48  & 0.05M & 0.0040 $\pm$ 0.0018 & 36.57 $\pm$ 2.56 \\
    Head      & 4   & 0.01M & 0.2776 $\pm$ 0.2312 & 19.01 $\pm$ 3.89 \\
    \bottomrule
  \end{tabular}
\end{table}

\Cref{tab:gradient_sel_vitb} presents the reconstruction results of gradient inversion attacks when only gradients from selected layers are utilized in the attacks. 
The "All" layers represent the default attack scenario, where gradients from all layers are employed.
The stem layers contain the patch embedding and position embedding processes.
These layers exhibit minimal changes in the output compared to the original image sample.
As the stem layers comprise only four layers, they can be relatively easier to attack compared to other types of layers. 
Consequently, the attacks on the stem layers demonstrate excellent performance, and we will further assess stem layers in the next subsection.

Among the attacks conducted with the remaining selected layers, the attack that demonstrates the best performance is the one utilizing "attention layers." 
This attack achieves an MSE of 0.0020 and a PSNR of 39.61. 
These results indicate that the attention layers are more susceptible to attacks, suggesting that they potentially leak more information about the data samples.

\begin{table}[t]
  \scriptsize
  \centering
  \caption{Attack performance on ConvNeXt-T compared with ResNet-50 and Swin-T. NN MIA for Network-based membership inference, AIA for attribute inference, and GIA for gradient inversion.}
    \resizebox{\linewidth}{!}{
    \begin{tabular}{c|c|ccc}
    \toprule
    \multicolumn{1}{r}{} &       & ResNet-50 & ConvNeXt-T & Swin-T \\
    \midrule
    \multicolumn{1}{l|}{NN MIA} & Attack acc $\uparrow$ & 0.6385 $\pm$ 0.0078 & 0.7471 $\pm$ 0.0052 & 0.6904 $\pm$ 0.0052 \\
    \midrule
    \multirow{2}[2]{*}{AIA} & Attack acc $\uparrow$  & 0.6854 $\pm$ 0.0015 & 0.7203 $\pm$ 0.0020 & 0.7312 $\pm$ 0.0014 \\
        & Macro F1 $\uparrow$ & 0.3753 $\pm$ 0.0012 & 0.5469 $\pm$ 0.0011 & 0.5530 $\pm$ 0.0019 \\
    \midrule
    \multirow{4}[2]{*}{GIA} & MSE $\downarrow$  & 1.5096 $\pm$ 0.5538 & 0.0177 $\pm$ 0.0171 & 0.0069 $\pm$ 0.0071 \\
          & PSNR $\uparrow$ & 10.58 $\pm$ 1.87 & 31.88 $\pm$ 5.04 & 36.24 $\pm$ 5.21 \\
          & LPIPS $\downarrow$ & 0.1624 $\pm$ 0.0613 & 0.0032 $\pm$ 0.0055 & 0.0012 $\pm$ 0.0016 \\
          & SSIM $\uparrow$ & 0.0896 $\pm$ 0.0544 & 0.9666 $\pm$ 0.0451 & 0.9892 $\pm$ 0.0118 \\
    \bottomrule
    \end{tabular}}%
  \label{tab:attack_convnextt}%
\end{table}%

\subsection{Impact of Other Micro Designs on Privacy}
\label{sec:impact_of_micro}

In the previous subsection, we established that attention layers within Transformers can contribute to privacy leakage. 
However, it is important to recognize that other micro designs within Transformers may also impact privacy vulnerabilities.
One example is ConvNeXt~\cite{liuConvNet2020s2022}, a convolutional neural network that incorporates multiple schemes from Transformers, similar to the Swin Transformer. 
ConvNeXt-T, ResNet-50, and Swin-T all share a similar parameter size, with ConvNeXt-T having approximately 27.83 million parameters. 
This allows for a direct comparison of the attack performance between ConvNeXt-T and the other two models.
When ConvNeXt-T is tested on CIFAR10 using the same attack settings, it achieves a task accuracy of 0.8258. 
This indicates that we can compare the attack performance of ConvNeXt-T with ResNet-50 and Swin-T.
The results presented in~\Cref{tab:attack_convnextt} further confirm the high attack performance on ConvNeXt-T. 
These findings suggest that ConvNeXt-T and Swin-T exhibit vulnerability to various attacks at a similar level.

The difference between ConvNeXt and the Swin Transformer lies in the attention module. 
This allows us to explore and investigate the privacy implications of various micro designs in model architectures beyond just the attention layers.
By studying the impact of different micro designs, we can gain deeper insights into the specific architectural features that may pose privacy risks.

We continue with utilizing gradient inversion attacks as a tool for further examination. 
Following the design process outlined in~\cite{liuConvNet2020s2022}, we have made several adjustments based on our own analysis.
Since ConvNeXt is constructed incrementally from ResNet, we meticulously scrutinized each model architecture to investigate the steps that contribute most significantly to privacy leakage. 
We focus on ResNet-50 and ConvNeXt-T models and conduct tests on a total of 14 model architectures using randomly selected samples from CIFAR10. 

The analysis involves a 14-step process, where we begin by modifying the overall architecture (Steps 1 to 4) based on ResNet-50. 
Subsequently, we align the bottleneck design (Steps 5 to 7) and refine the stem layer design in Step 8. 
Finally, we align the micro designs (Steps 9 to 14) to achieve ConvNeXt-T.
This 14-step process, from macro to micro levels, as outlined in~\cite{liuConvNet2020s2022}, has been established as an optimal procedure for designing ConvNeXt. 
Leveraging this proven methodology, we utilize it to analyze the impact on privacy.
By identifying the critical steps that contribute to privacy leakage, we can assess the influence of micro designs on the privacy of model architectures. 
This analysis provides valuable insights into understanding the specific stages or modifications that may pose risks to privacy.
The 14 steps are outlined below:

\begin{enumerate}
\itemsep-0.28em 
\item \textbf{ResNet-50:} 
We begin our process with this model.
\item \textbf{Changing channel dimensions:} 
Each stage in ResNet-50 uses different channel dimensions (i.e., $(64, 128, 256, 512)$). 
To align with ConvNeXt-T, we modify these dimensions to $(96, 192, 384, 768)$.
\item \textbf{Changing the stage compute ratio:} 
ResNet employs a multi-stage design that modifies channel dimensions. 
ResNet-50 has a stage compute ratio of $(3, 4, 6, 3)$, while ConvNeXt and Swin Transformers adopt $(3, 3, 9, 3)$. 
We follow this adjustment in this step.
\item \textbf{Applying "Patchify":} 
Vision transformers process input images by sliding them into patches. 
Here, we replace the stem convolutional layers with a kernel size of $(4 \times 4)$ and a stride of 4.
\item \textbf{Applying "ResNeXtify":} 
ResNeXt~\cite{xieAggregatedResidualTransformations2017} introduced grouped convolution, reducing parameter size while maintaining performance. 
We utilize depth-wise convolution, which employs the same number of groups and channels.
\item \textbf{Using the inverted bottleneck:} 
The inverted bottleneck design is widely employed in models like MobileNet, ConvNeXt, and Swin Transformers. 
We incorporate this step into our process.
\item \textbf{Enlarging kernel sizes:} 
To align the parameters with ConvNeXt, we adopt a larger kernel size of $(7 \times 7)$ instead of $(3 \times 3)$.
\item \textbf{Forming the new stem layers:} 
In this step, we remove the activation and maxpool layers, originally part of ResNet.
\item \textbf{Changing ReLU to GELU:} 
The Gaussian Error Linear Unit (GELU)~\cite{hendrycksGaussianErrorLinear2016} is a variant of ReLU commonly used in Transformers. 
We introduce this change to the model.
\item \textbf{Removing some activation layers:} 
Transformer blocks typically have fewer activation layers. 
Here, we retain only one activation layer within the block.
\item \textbf{Removing some normalization layers:} 
We reduce the number of BatchNorm (BN) layers within the block to one.
\item \textbf{Changing BN to LN:} 
Inspired by the prevalent use of LayerNorm (LN) layers~\cite{baLayerNormalization2016} in Transformers, we replace the BN layers in our model with LN layers.
We also enable bias parameters for all convolutional layers in this step.
\item \textbf{Separating downsampling layers:} 
We move the downsampling layers between stages, introducing an LN layer to ensure stability during training.
\item \textbf{Final touches to reach ConvNeXt-T:} 
We incorporate Stochastic Depth~\cite{huangDeepNetworksStochastic2016} and Layer Scale~\cite{touvronGoingDeeperImage2021} in the final stage to complete the ConvNeXt-T model.
\end{enumerate}

Please refer to~\cite{zhangHowDoesDeep2024} for more detailed architecture specifications.

\begin{table*}[t]
  \scriptsize
  \caption{The results of gradient inversion attacks on model architectures from 14 steps. Some significant changes in results are marked in bold.}
  \label{tab:arch_name_and_result}
  \centering
  \begin{tabular}{lccccc}
  \toprule
  Steps & Task acc $\uparrow$ & MSE $\downarrow$ & PSNR $\uparrow$ & LPIPS $\downarrow$ & \textbf{SSIM $\uparrow$} \\
  \midrule
  1. ResNet-50      & 0.8220 $\pm$ 0.0039 & 1.5096 $\pm$ 0.5538 & 10.58 $\pm$ 1.87  & 0.1624 $\pm$ 0.0613 & 0.0896 $\pm$ 0.0544 \\
  2. Channel dim    & 0.8240 $\pm$ 0.0072 & 1.4706 $\pm$ 0.5710 & 10.74 $\pm$ 1.97  & 0.1724 $\pm$ 0.0616 & 0.0826 $\pm$ 0.0405 \\
  3. Stage ratio    & 0.8282 $\pm$ 0.0040 & 1.5286 $\pm$ 0.5246 & 10.56 $\pm$ 2.05  & 0.1834 $\pm$ 0.0581 & 0.0731 $\pm$ 0.0613 \\
  4. Patchify       & 0.8293 $\pm$ 0.0061 & \textbf{0.9011 $\pm$ 0.4376} & \textbf{12.97 $\pm$ 2.10} & \textbf{0.0867 $\pm$ 0.0436} & \textbf{0.1727 $\pm$ 0.0794} \\
  5. ResNeXtify     & 0.8397 $\pm$ 0.0033 & 1.2415 $\pm$ 0.6934 & 11.86 $\pm$ 2.77 & 0.1066 $\pm$ 0.0391 & 0.1334 $\pm$ 0.0950 \\
  6. Inv bottleneck & 0.8407 $\pm$ 0.0058 & 1.1123 $\pm$ 0.4994 & 12.06 $\pm$ 2.19 & 0.0989 $\pm$ 0.0290 & 0.1429 $\pm$ 0.0844 \\
  7. Kernel sizes   & 0.8432 $\pm$ 0.0052 & 0.8206 $\pm$ 0.3543 & 13.40 $\pm$ 2.30 & 0.0821 $\pm$ 0.0355 & 0.2353 $\pm$ 0.0766 \\
  8. New stem       & 0.8459 $\pm$ 0.0043 & 0.5684 $\pm$ 0.3564 & 15.43 $\pm$ 3.01 & 0.0752 $\pm$ 0.0381 & 0.4924 $\pm$ 0.1205 \\
  9. ReLU to GELU   & 0.8436 $\pm$ 0.0027 & 1.0540 $\pm$ 0.5075 & 12.42 $\pm$ 2.61 & 0.2422 $\pm$ 0.0904 & 0.1746 $\pm$ 0.1166 \\
  10. Removing Act  & 0.8480 $\pm$ 0.0064 & \textbf{0.0215 $\pm$ 0.0150} & \textbf{29.93 $\pm$ 3.58} & \textbf{0.0049 $\pm$ 0.0026} & \textbf{0.9562 $\pm$ 0.0224} \\
  11. Removing BN   & 0.8491 $\pm$ 0.0059 & 0.0198 $\pm$ 0.0139 & 30.57 $\pm$ 4.12 & 0.0045 $\pm$ 0.0032 & 0.9605 $\pm$ 0.0232 \\
  12. BN to LN      & 0.8501 $\pm$ 0.0031 & \textbf{0.0049 $\pm$ 0.0044} & \textbf{36.86 $\pm$ 3.96} & \textbf{0.0005 $\pm$ 0.0003} & \textbf{0.9927 $\pm$ 0.0064} \\
  13. Sep downsamp  & 0.8553 $\pm$ 0.0070 & 0.0121 $\pm$ 0.0171 & 33.79 $\pm$ 4.69 & 0.0011 $\pm$ 0.0008 & 0.9859 $\pm$ 0.0151 \\
  14. ConvNeXt      & 0.8523 $\pm$ 0.0064 & 0.0177 $\pm$ 0.0171 & 31.88 $\pm$ 5.04 & 0.0032 $\pm$ 0.0055 & 0.9666 $\pm$ 0.0451 \\
  \bottomrule
  \end{tabular}
\end{table*}

\begin{figure}[t]
    \scriptsize
    \centering
    \includegraphics[width=1\linewidth]{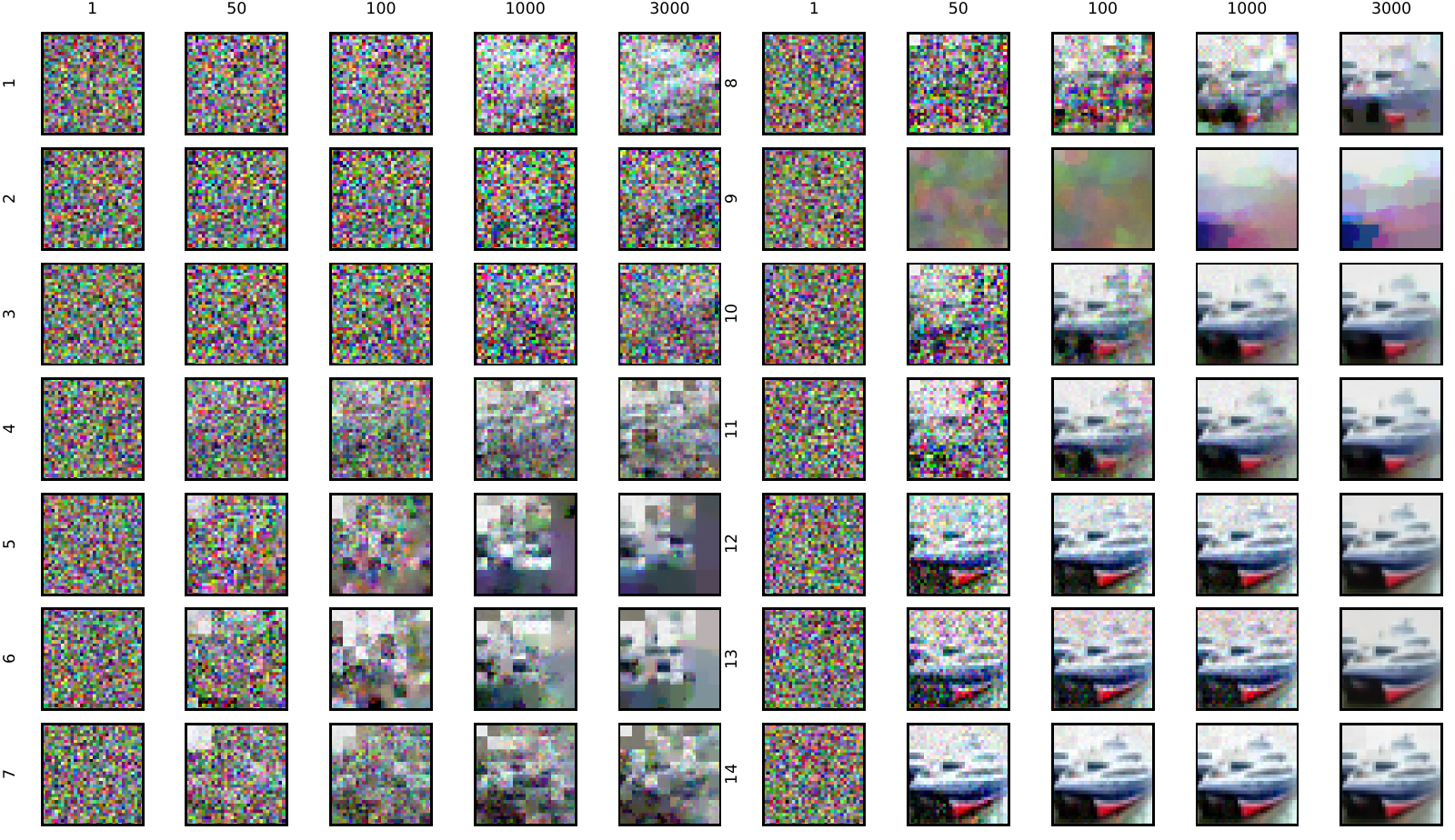}
    \caption{The performance of gradient inversion attacks on each architecture changing from ResNet-50 to ConvNeXt-T with several selected iterations (i.e., 1, 50, 100, 1000, 3000). Model architectures from 14 steps are shown.}
    \label{fig:arch_change_img_multi_iter}
\end{figure}

\Cref{fig:arch_change_img_multi_iter} and~\Cref{tab:arch_name_and_result} show the performance of gradient inversion attacks on each model architecture.
The 14-step changes from macro to micro levels exhibit fluctuations but generally show a trend of improved attack performance.

\Cref{fig:arch_change_img_multi_iter} provides qualitative results of the attacks.
During the initial stages (Steps 1 to 3), the attacks struggle to reconstruct proper images.
In the middle stages, some information emerges in the reconstruction results, but to a limited extent.
The reconstruction results improve in the later stages (After Step 10).
At last, using ConvNeXt-T, which is step 14, the attacks achieve a good attack performance.

\Cref{tab:arch_name_and_result} presents more information on gradient inversion attacks, highlighting their performance in various architectural changes. 
Several steps exhibit significant changes compared to other steps.
We define a step as a significant change when a metric notably decreases or increases. 
To illustrate, we consider the MSE metric, which we define a step as a significant change if it experiences a reduction exceeding $50\%$.
Our analysis reveals three significant increases in attack performance, each correlated with specific architectural changes:
The first one occurs when applying "Patchify" to stem layers;
The second one occurs with the removal of some activation layers;
The third one happens when changing BN to LN.
These findings underscore the pivotal role these specific architectural changes play in determining the model's vulnerability to privacy attacks.

\subsection{Ablation Study on Micro Designs}
\label{sec:ablation_study}

\begin{figure}[t]
  \scriptsize
  \centering
  \includegraphics[width=1\linewidth]{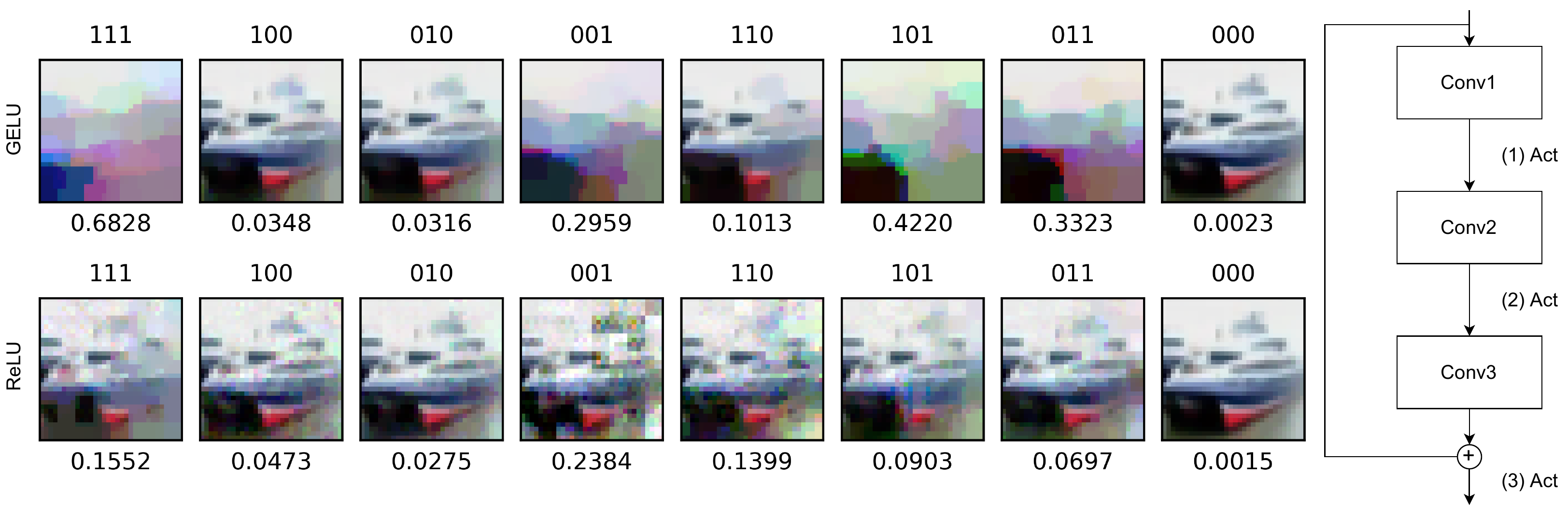}
  \caption{The performance of gradient inversion attacks on models with different positions of activation layers. The position of three activation layers is illustrated on the right, with two between convolutional layers and one after the addition operation. The three digits on the top of the subfigures show whether the activation layer on this position is added or not. The bottom of each subfigure provides MSE values for the attack on this model (i.e., models with GELU or ReLU).}
  \label{fig:act_layer}
\end{figure}

\begin{figure}[t]
  \scriptsize
  \begin{subfigure}[t]{.23\textwidth}
    \centering
    \includegraphics[width=\linewidth]{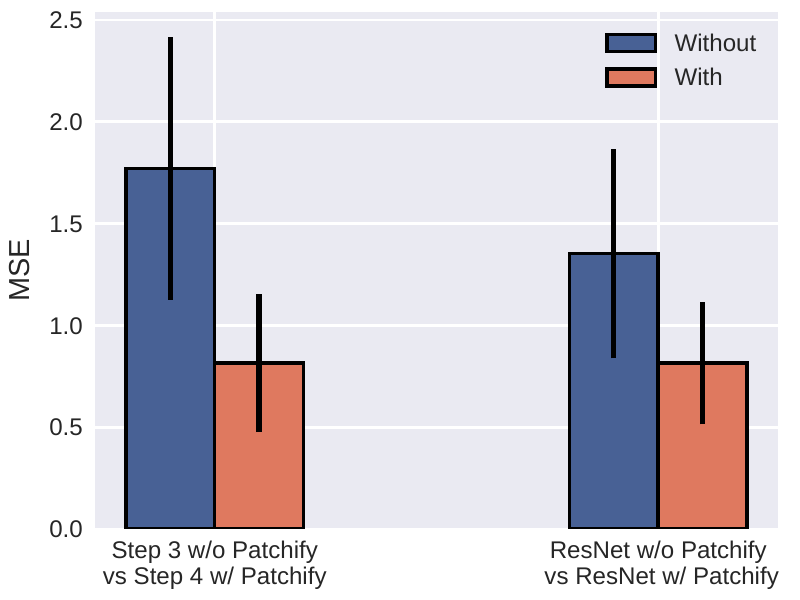}
    \caption{Patchify}
    \label{fig:patchify}
  \end{subfigure}
  \hfill
  \begin{subfigure}[t]{.23\textwidth}
    \centering
    \includegraphics[width=\linewidth]{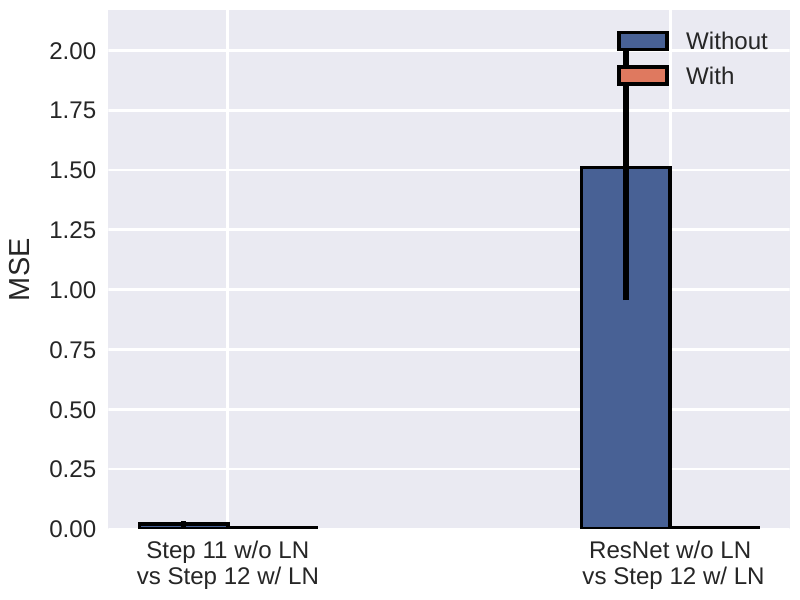}
    \caption{LN}
    \label{fig:ln}
  \end{subfigure}
  \caption{The performance of gradient inversion attacks on model architectures with or without some micro designs.}
  \label{fig:arch_feature_analysis}
\end{figure}

\textbf{Ablation study on the design of activation layers.}
One of the differences between a Transformer block and a ResNet block is that a Transformer block has fewer activation layers.
Leaving fewer activation layers in the model boosts the attack performance.
As illustrated in~\Cref{fig:act_layer}, when different activation layers are removed, there is a noticeable improvement in the attack performance. 
Particularly, the removal of the third activation layer, located after the skip connection of the ResNet block, results in a significant enhancement in attack accuracy.
This observation suggests that this activation layer introduces a non-linear process that reduces the amount of information available for the attack, thus making it harder to reconstruct the original input. 
By removing this layer, the attack performance is improved.
The analysis further highlights that changing ReLU to GELU results in a marginal reduction in attack performance. 
This is attributed to the fact that GELU is a smoother approximation of ReLU, which strengthens the model's robustness and generalization~\cite{baiAreTransformersMore2021}.
Using GELU makes the adversary extract less information from data samples, consequently decreasing their attack efficacy. 
In sum, the most significant enhancement in attack performance occurs when all activation layers are removed from the model architecture.

\textbf{Ablation study on the design of stem layers and LN layers.}
The additional analysis presented in~\Cref{fig:arch_feature_analysis} highlights two more features that appear to have an impact on model privacy leakage.
Firstly,~\Cref{fig:patchify} demonstrates that the utilization of "Patchify" reduces reconstruction MSE results. 
As this process involves modifications to the stem layers, we believe that these stem layers are crucial for privacy attacks. 
Secondly,~\Cref{fig:ln} illustrates that changing BN to LN results in improved reconstruction results. 
This suggests that incorporating LN also contributes to the model's privacy leakage.
These findings emphasize the significance of stem layers and LN layers on model privacy leakage. 

\begin{figure}[t]
  \scriptsize
  \begin{subfigure}[t]{.45\textwidth}
    \centering
    \includegraphics[width=\linewidth]{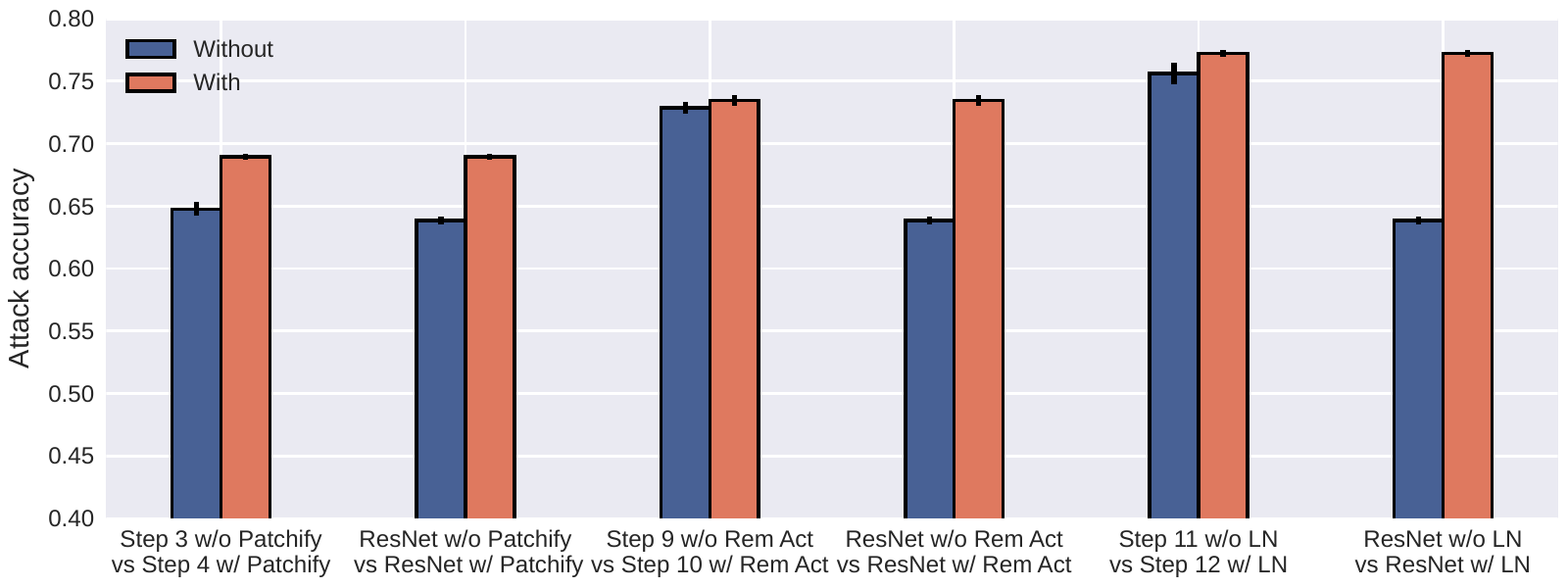}
    \caption{Membership inference}
    \label{fig:micro_design_mia}
  \end{subfigure}
  \vfill
  \begin{subfigure}[t]{.45\textwidth}
    \centering
    \includegraphics[width=\linewidth]{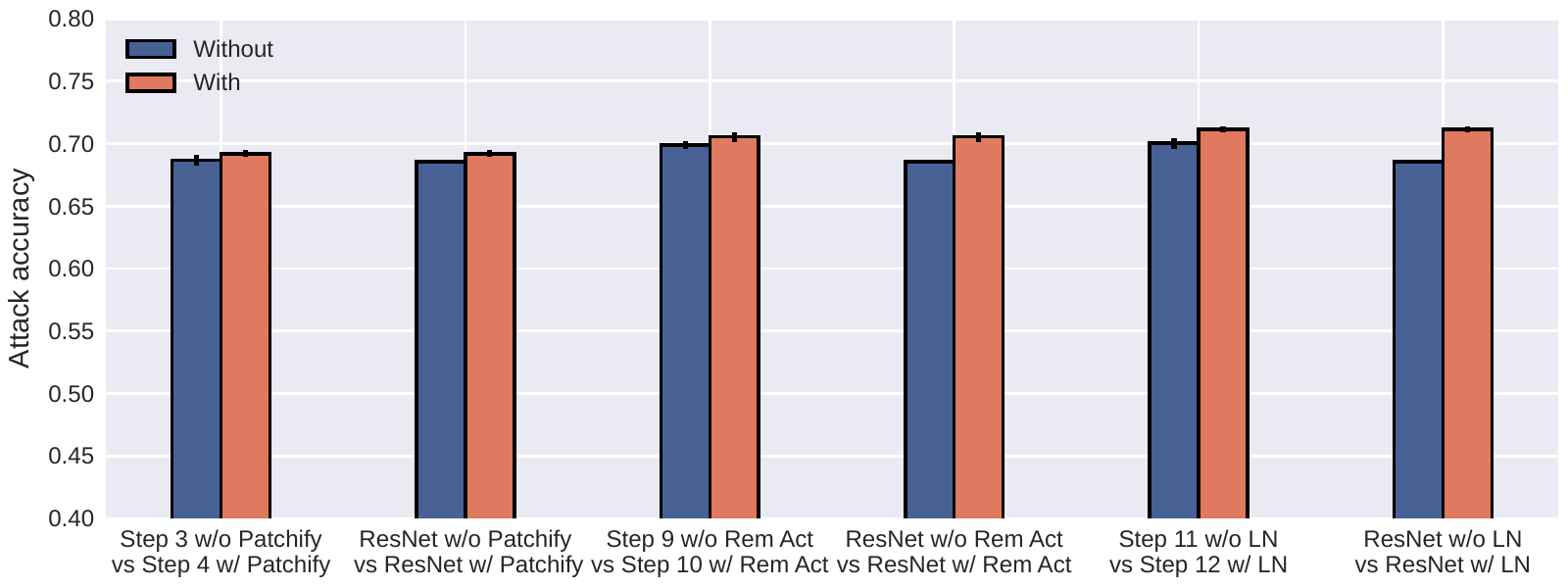}
    \caption{Attribute inference}
    \label{fig:micro_design_aia}
  \end{subfigure}
  \caption{The performance of privacy attacks on model architectures with or without some features.}
  \label{fig:micro_design_mia_and_aia}
\end{figure}

\textbf{Ablation study on micro designs for membership inference attacks and attribute inference attacks.}
\Cref{fig:micro_design_mia_and_aia} shows the impact of three micro designs on more attacks.
Just like previous studies, adding these designs can also increase privacy leakage in these two attacks.

\textbf{In summary}, attention modules, as well as the design of activation layers, stem layers, and LN layers, are the key architectural features that lead to more privacy leakage.
We provide more discussions in~\Cref{sec:discussion}.

\section{Discussion}
\label{sec:discussion}

In the previous sections, 
we discovered that four design components in Transformers could result in privacy leakage: attention modules, activation layers, stem layers, and LN layers.
In this section, 
we would like to provide a more in-depth discussion of these modules.

\subsection{The Impact of Attention Modules}
\label{sec:discussion_attention}

The receptive field of a model refers to the information received within a specified range by a neuron in a model layer.
In a fully connected neural network, 
each neuron receives input from the elements of the entire input sample.
Due to the convolution operation, 
the neuron in a convolutional network receives input limited to its receptive field.
The range of the receptive field is defined by the convolution templates in CNNs.
This design allows CNNs to capture local patterns in the input data efficiently.
The receptive field has a theoretical limit.
Some researchers have demonstrated that the effective receptive field (i.e., the receptive field's effective area) is smaller than the theoretical receptive field~\cite{luoUnderstandingEffectiveReceptive2016}.
From a privacy perspective, 
a CNN model could only reveal part of sensitive information from the input sample due to the design of localized convolution templates.

Transformers employ the multi-head self-attention mechanism, 
also known as attention modules.
The input sample is taken as a sequence of flattened 2D patches.
The attention module receives the input sequence and generates its representation of the sequence by mapping the query and the key-value pairs to the output.
Transformers tend to have much larger receptive fields than CNNs due to the fact that their attention module is computed with the entire input sequence~\cite{vaswaniAttentionAllYou2017,dosovitskiyImageWorth16x162021}.
In terms of privacy, a Transformer model is able to extract more sensitive information than a CNN model because of its wide-angle receptive field.
Hence, 
Transformers are more prone to attacks than CNNs, 
as demonstrated by our evaluation based on three popular privacy attacks.

\subsection{The Impact of Micro Designs}
\label{sec:discussion_micro}

There are micro-design components with the potential to leak sensitive data from input samples.
Activation layers such as ReLU and GELU add a layer of complexity to the model by making it a non-linear function.
Removing some activation layers simplifies the logic of attacks.
Stem layers receive an input sample and perform some preliminary processing.
As the representation after the stem layers remains similar to the original input image, 
there is a high possibility of extracting private information because of the design of the stem layers.
Changing BN to LN is also likely to aid the attack process and allow the adversary to achieve higher attack performance.

\subsection{The Impact of Overfitting}
\label{sec:discussion_overfitting}

Previous work claimed that privacy attacks are caused mainly by the undesirable overfitting issue in deep learning models~\cite{shokriMembershipInferenceAttacks2017,liuMLDoctorHolisticRisk2022}.
Overfitting normally occurs when a model performs well on the training data, 
but poorly on the validation data.
The overfitting issue tends to become severe on an over-trained model with a large number of parameters.
Deep learning models are exposed to privacy threats due to the overfitting effect.
In our work, 
we find that model architectures have impacts on the performance of privacy attacks, 
which can \emph{not} be attributed solely to the overfitting effect.
Indeed, 
our experiments validate that the variation in performance is due to the difference in model architectures. 
For models with the same level of parameter sizes, 
Transformers tend to be more vulnerable to privacy attacks than CNNs. 
More importantly, 
for models with the same overfitting level, 
our conclusion still holds that Transformers are more vulnerable to privacy attacks than CNNs. 
We have then identified some architectural features that could be responsible for privacy leakage.

\subsection{Potential (Incorrect) Explanations for the Vulnerability of Different Models Against Privacy Attacks}
\label{sec:discussion_potential}

Here, we explore several potential explanations for our experimental results and shed more insights on our conclusion.
\begin{itemize}
\itemsep-0.28em 
    \item The attacks on Transformers are more effective than those on CNNs. Is it due to random noise? 
    To dismiss this concern, 
    we have conducted multiple runs for each experiment setting and calculated the mean and standard deviation scores for our results. 
    In total, 
    We have done around 100 different experiment settings, 
    involving 1400 experiments, 1700 training models, and 1200 training hours, 
    which averages out the impact of random noise.
    
    \item Is it due to the overfitting of the victim models? 
    To account for this, 
    we meticulously compare the attacks on CNNs and Transformers when their victim models are trained on the same level of overfitting. 
    By doing so, 
    we minimize the factor of overfitting for a fair assessment.
    
    \item Is it due to the immature training of the victim models? 
    In order to conduct our experiments, 
    it was necessary to split the dataset into multiple subsets for both victim and shadow models. 
    Consequently, 
    it is reasonable to expect relatively low accuracy on the victim models for CIFAR10, CIFAR100, and CelebA (compared to the models trained on a full dataset). 
    This aligns with the findings of the literature on privacy attacks, 
    which have reported similar results~\cite{heQuantifyingMitigatingPrivacy2021,carliniMembershipInferenceAttacks2022,liuMLDoctorHolisticRisk2022}.
    
    \item Is it due to other training factors? 
    We ensure that the training recipes for each victim model remain consistent across all comparisons. 
    Consequently, 
    we can confidently exclude other training factors from the comparison study when drawing our conclusion.
\end{itemize}

After eliminating these potential explanations, 
we believe that our experimental results should be best explained by several architectural features mentioned in~\Cref{sec:which_arch}.
Please see~\Cref{sec:more_theoretical_discussion} for more theoretical discussions.

\begin{figure}[t]
  \scriptsize
  \begin{subfigure}[t]{.23\textwidth}
    \centering
    \includegraphics[width=\linewidth]{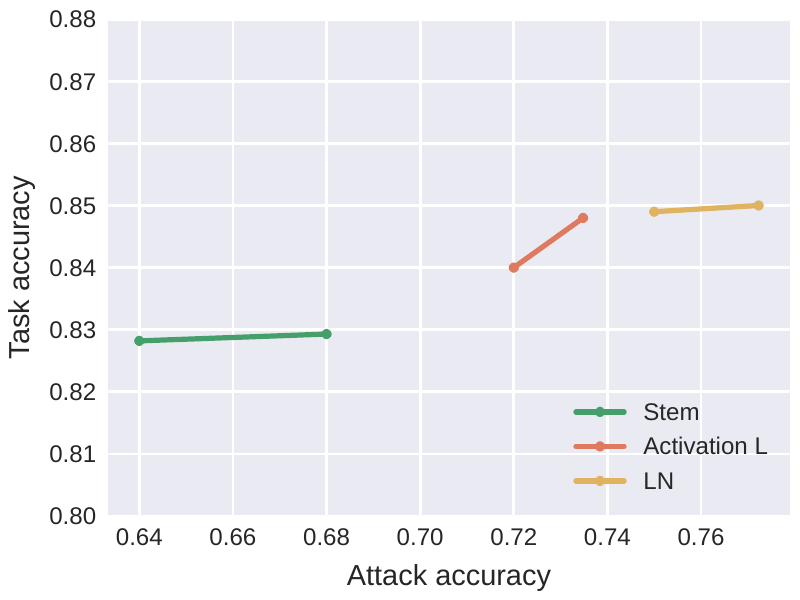}
    \caption{Membership inference}
    \label{fig:defense_mia}
  \end{subfigure}
  \hfill
  \begin{subfigure}[t]{.23\textwidth}
    \centering
    \includegraphics[width=\linewidth]{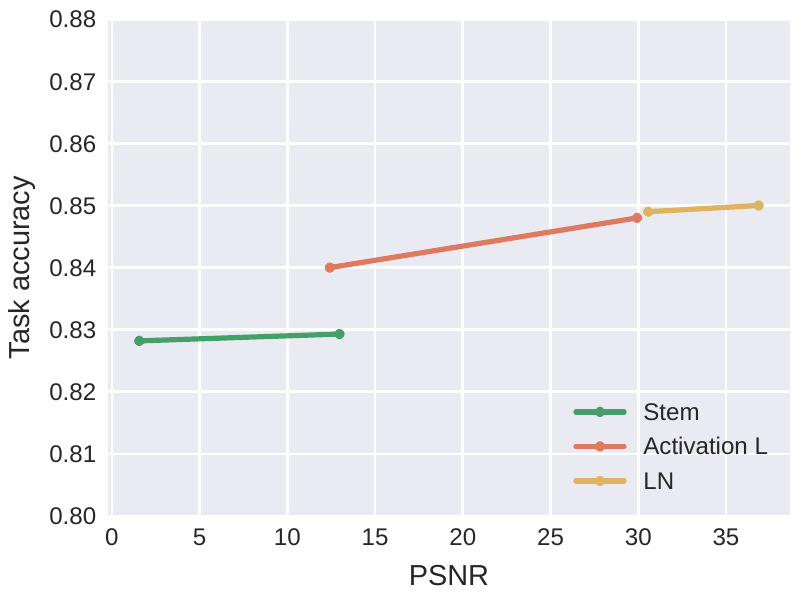}
    \caption{Gradient inversion}
    \label{fig:defense_gia}
  \end{subfigure}
  \caption{The impact of the performance of privacy attacks when model components change. In each line, the left and right points represent the results without and with the micro design modifications, respectively. }
  \label{fig:defense}
\end{figure}

\section{How to Exploit the Privacy Impact of Model Components?}
\label{sec:defense}

\subsection{Modifying Model Components as a Defense Mechanism}
\label{sec:defense_modifying}

Based on our discoveries in previous sections, 
we propose to explore the modification of model components as a defense mechanism against privacy attacks. 
We aim to achieve a trade-off between utility and privacy.
In~\Cref{fig:defense}, 
we demonstrate the influence of certain micro-design changes on the efficacy of privacy attacks.
Notably, 
we observe that the task accuracy is minimally impacted in each test, 
indicating that the proposed defense measures do not significantly compromise the model's overall performance. 
However, 
we can observe a decreased effectiveness of membership inference attacks and gradient inversion attacks when these model components are not applied.

These findings suggest that the proposed micro-design modifications can serve as effective countermeasures against privacy attacks,  
while the task accuracy remains almost intact.

\begin{figure}[t]
  \scriptsize
  \centering
  \includegraphics[width=1\linewidth]{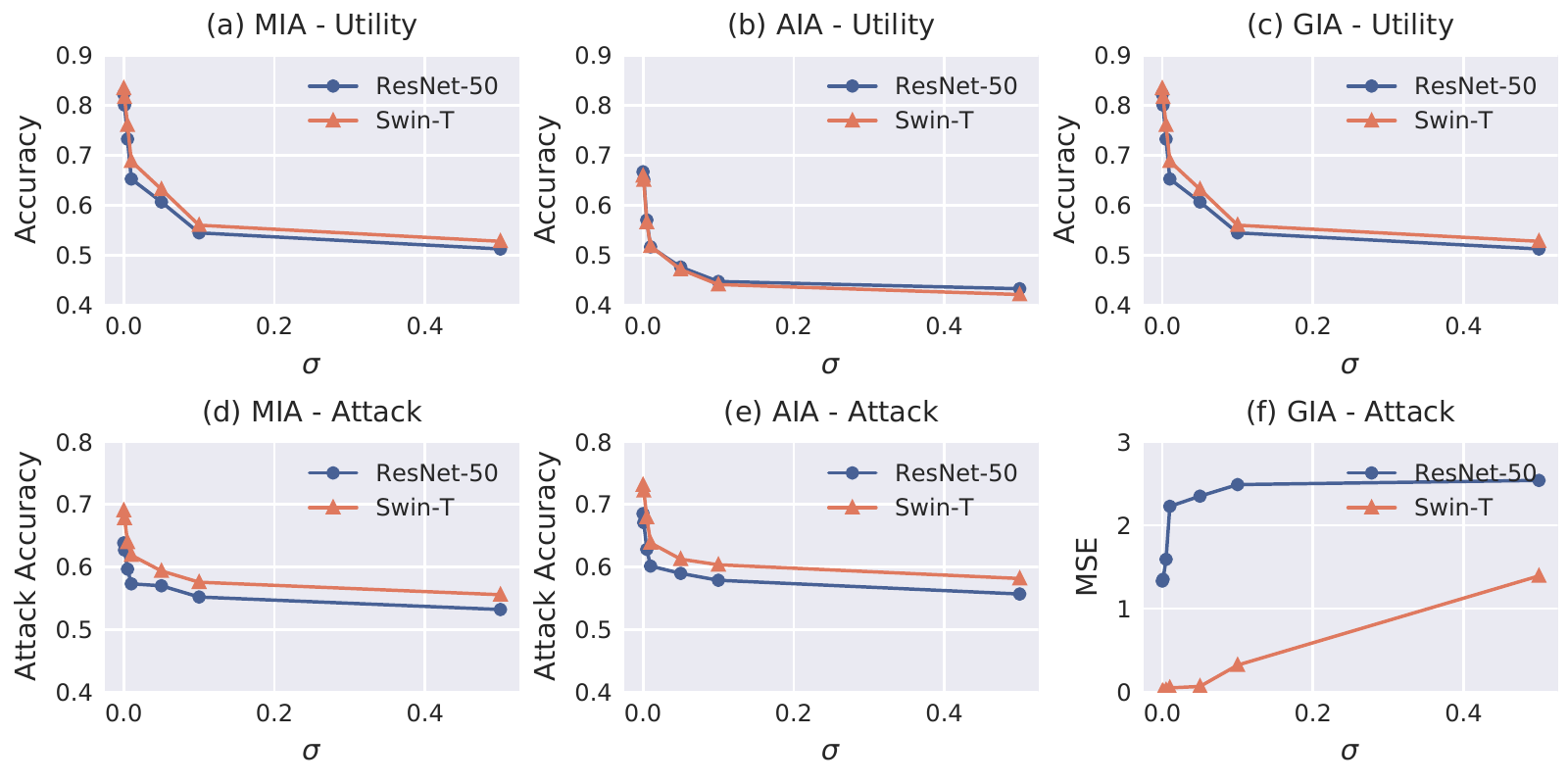}
  \caption{The performance of privacy attacks when DPSGD is applied. The utility performance of victim models and the attack performances are given when various DP noises are given.}
  \label{fig:dp}
\end{figure}

\begin{figure}[t]
  \scriptsize
  \centering
  \includegraphics[width=1\linewidth]{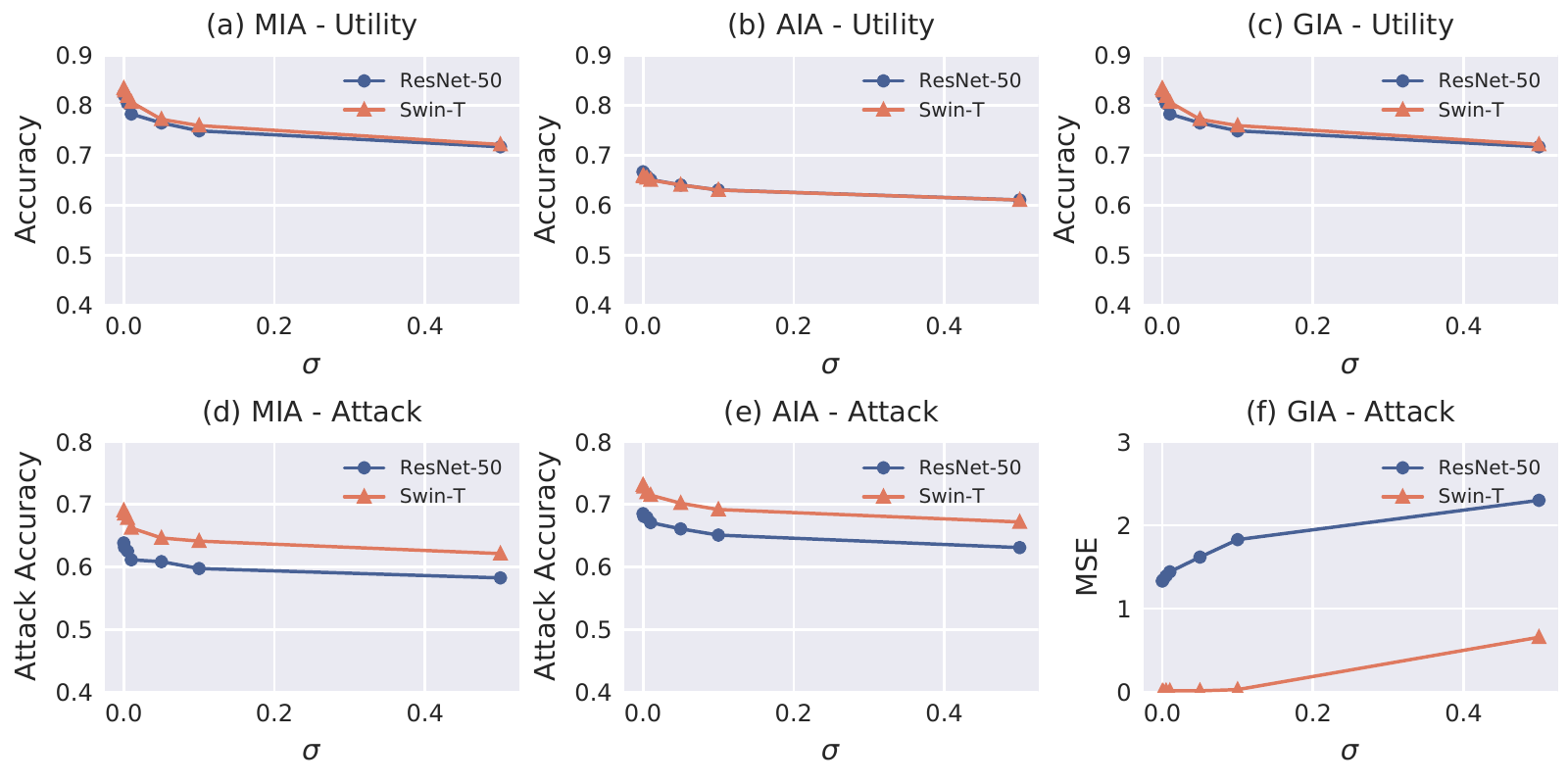}
  \caption{The performance of privacy attacks when DPSGD is applied in stem layers. The utility performance of victim models and the attack performances are given when various DP noises are given.}
  \label{fig:dp_stem_layer}
\end{figure}

\subsection{Adding Perturbations as a Defense Mechanism}
\label{sec:defense_perturbation}

The privacy leakage issue observed in Transformers highlights the need for an enhanced privacy treatment compared to CNNs. 
Specifically, 
when employing perturbation as a defense mechanism, such as incorporating differential privacy noises into the model parameters, 
it is better to increase the level of noise specifically for Transformers and Transformer-like models.
\Cref{fig:dp} illustrates the outcomes of three privacy attacks when DPSGD~\cite{abadiDeepLearningDifferential2016} is applied.
As the differential privacy noise increases, a noticeable decline in the utility performance of victim models is observed (depicted in subfigures (a) to (c)).
Concurrently, subfigures (d) to (f) reveal diminishing attack performances, with Transformers (Swin-T) consistently exhibiting superior attack performance at equivalent noise levels. 
From a defensive standpoint, adding more noise to Transformers is recommended.

Another approach to consider is a layer-wise perturbation defense mechanism, 
where noises are added to selected layers only. 
In this scenario, 
it would be interesting to introduce additional noises to the layers that are more susceptible to privacy leakages, 
such as activation layers, stem layers, LN layers, and attention modules. 
Paying special attention to these "privacy-leakage" layers could also achieve satisfactory privacy protection.
\Cref{fig:dp_stem_layer} demonstrates the impact of privacy attacks when DPSGD is applied to stem layers.
As the differential privacy noise escalates, it influences both utility and attack performance, demonstrating the efficacy of adding noises to targeted layers.

\section{Conclusion}

In this study, we pioneer the exploration of privacy risks on model architectures, especially CNNs and Transformers.
We have conducted a comparison of three prominent privacy attacks, 
i.e., 
membership inference attacks, 
attribute inference attacks, 
and gradient inversion attacks.
We discover that Transformers tend to be more vulnerable to privacy attacks than CNNs.
As a result, 
many Transformers-inspired CNN designs, 
such as ConvNeXt, 
are also susceptible to privacy threats.
A number of Transformers' features,
including the design of activation layers, 
the design of stem layers, 
the design of LN layers,
and the attention modules,
are likely to incur high privacy risks.

It is still challenging to establish accurate and theoretical explanations for why certain architectural features are critical to privacy preservation.
We believe that these analyses require further experimental campaigns, 
and we intend to study this matter in our future work.

\section*{Acknowledgments}

We thank all anonymous reviewers and our shepherd for their constructive comments. 
This work is funded in part by ARC Discovery Grant (DP230100246) and ARC Linkage Grant (LP220200808).

{\footnotesize
\bibliographystyle{plain}
\bibliography{bib}

\begin{thebibliography}{10}

\bibitem{abadiDeepLearningDifferential2016}
Martin Abadi, Andy Chu, Ian Goodfellow, H.~Brendan McMahan, and et~al.
\newblock Deep {{Learning}} with {{Differential Privacy}}.
\newblock In {\em {{CCS}}}, 2016.

\bibitem{araujoComputingReceptiveFields2019}
Andr{\'e} Araujo, Wade Norris, and Jack Sim.
\newblock Computing receptive fields of convolutional neural networks.
\newblock {\em Distill}, 2019.

\bibitem{baLayerNormalization2016}
Jimmy~Lei Ba, Jamie~Ryan Kiros, and Geoffrey~E. Hinton.
\newblock Layer {{Normalization}}.
\newblock {\em arXiv preprint arXiv:1607.06450}, 2016.

\bibitem{baiAreTransformersMore2021}
Yutong Bai, Jieru Mei, Alan~L Yuille, and Cihang Xie.
\newblock Are {{Transformers}} more robust than {{CNNs}}?
\newblock In {\em {{NeurIPS}}}, 2021.

\bibitem{brownLanguageModelsAre2020}
Tom Brown, Benjamin Mann, Nick Ryder, Melanie Subbiah, and et~al.
\newblock Language {{Models}} are {{Few-Shot Learners}}.
\newblock In {\em {{NeurIPS}}}, 2020.

\bibitem{bubeckNetworkSizeSize2020}
Sebastien Bubeck, Ronen Eldan, Yin~Tat Lee, and Dan Mikulincer.
\newblock Network size and size of the weights in memorization with two-layers
  neural networks.
\newblock In {\em {{NeurIPS}}}, 2020.

\bibitem{carliniMembershipInferenceAttacks2022}
Nicholas Carlini, Steve Chien, Milad Nasr, Shuang Song, Andreas Terzis, and
  Florian Tram{\`e}r.
\newblock Membership {{Inference Attacks From First Principles}}.
\newblock In {\em {{S\&P}}}, 2022.

\bibitem{carliniPrivacyOnionEffect2022}
Nicholas Carlini, Matthew Jagielski, Chiyuan Zhang, Nicolas Papernot, Andreas
  Terzis, and Florian Tramer.
\newblock The {{Privacy Onion Effect}}: {{Memorization}} is {{Relative}}.
\newblock In {\em {{NeurIPS}}}, 2022.

\bibitem{chenGANLeaksTaxonomyMembership2020}
Dingfan Chen, Ning Yu, Yang Zhang, and Mario Fritz.
\newblock {{GAN-Leaks}}: {{A Taxonomy}} of {{Membership Inference Attacks}}
  against {{Generative Models}}.
\newblock In {\em {{CCS}}}, 2020.

\bibitem{chenWhenMachineUnlearning2021}
Min Chen, Zhikun Zhang, Tianhao Wang, Michael Backes, Mathias Humbert, and Yang
  Zhang.
\newblock When {{Machine Unlearning Jeopardizes Privacy}}.
\newblock In {\em {{CCS}}}, 2021.

\bibitem{choquette-chooLabelOnlyMembershipInference2021}
Christopher~A. {Choquette-Choo}, Florian Tramer, Nicholas Carlini, and Nicolas
  Papernot.
\newblock Label-{{Only Membership Inference Attacks}}.
\newblock In {\em {{ICML}}}, 2021.

\bibitem{collinsCapacityTrainabilityRecurrent2017}
Jasmine Collins, Jascha {Sohl-Dickstein}, and David Sussillo.
\newblock Capacity and {{Trainability}} in {{Recurrent Neural Networks}}.
\newblock In {\em {{ICLR}}}, 2017.

\bibitem{dengImageNetLargescaleHierarchical2009}
Jia Deng, Wei Dong, Richard Socher, Li-Jia Li, Kai Li, and Li~{Fei-Fei}.
\newblock {{ImageNet}}: {{A}} large-scale hierarchical image database.
\newblock In {\em {{CVPR}}}, 2009.

\bibitem{devlinBERTPretrainingDeep2019}
Jacob Devlin, Ming-Wei Chang, Kenton Lee, and Kristina Toutanova.
\newblock {{BERT}}: {{Pre-training}} of {{Deep Bidirectional Transformers}} for
  {{Language Understanding}}.
\newblock In {\em {{NAACL-HLT}}}, 2019.

\bibitem{dingDaViTDualAttention2022}
Mingyu Ding, Bin Xiao, Noel Codella, Ping Luo, Jingdong Wang, and Lu~Yuan.
\newblock Davit: Dual attention vision transformers.
\newblock In {\em {{ECCV}}}, 2022.

\bibitem{dosovitskiyImageWorth16x162021}
Alexey Dosovitskiy, Lucas Beyer, Alexander Kolesnikov, Dirk Weissenborn,
  Xiaohua Zhai, and et~al.
\newblock An {{Image}} is {{Worth}} 16x16 {{Words}}: {{Transformers}} for
  {{Image Recognition}} at {{Scale}}.
\newblock In {\em {{ICLR}}}, 2021.

\bibitem{duWhenConvolutionalFilter2018}
Simon~S. Du, Jason~D. Lee, and Yuandong Tian.
\newblock When is a {{Convolutional Filter Easy}} to {{Learn}}?
\newblock In {\em {{ICLR}}}, 2018.

\bibitem{dudduInferringSensitiveAttributes2022}
Vasisht Duddu and Antoine Boutet.
\newblock Inferring {{Sensitive Attributes}} from {{Model Explanations}}.
\newblock In {\em {{CIKM}}}, 2022.

\bibitem{feldmanDoesLearningRequire2020}
Vitaly Feldman.
\newblock Does learning require memorization? a short tale about a long tail.
\newblock In {\em {{STOC}}}, 2020.

\bibitem{fredriksonModelInversionAttacks2015}
Matt Fredrikson, Somesh Jha, and Thomas Ristenpart.
\newblock Model {{Inversion Attacks}} that {{Exploit Confidence Information}}
  and {{Basic Countermeasures}}.
\newblock In {\em {{CCS}}}, 2015.

\bibitem{fredriksonPrivacyPharmacogeneticsEndtoEnd2014}
Matthew Fredrikson, Eric Lantz, Somesh Jha, Simon Lin, David Page, and et~al.
\newblock Privacy in {{Pharmacogenetics}}: {{An End-to-End Case Study}} of
  {{Personalized Warfarin Dosing}}.
\newblock In {\em {{USENIX Security}}}, 2014.

\bibitem{geipingInvertingGradientsHow2020}
Jonas Geiping, Hartmut Bauermeister, Hannah Dr{\"o}ge, and Michael Moeller.
\newblock Inverting {{Gradients}} - {{How}} easy is it to break privacy in
  federated learning?
\newblock {\em {{NeurIPS}}}, 2020.

\bibitem{goelLearningOneConvolutional2018}
Surbhi Goel, Adam Klivans, and Raghu Meka.
\newblock Learning {{One Convolutional Layer}} with {{Overlapping Patches}}.
\newblock In {\em {{ICML}}}, 2018.

\bibitem{hatamizadehGradViTGradientInversion2022}
Ali Hatamizadeh, Hongxu Yin, Holger~R. Roth, Wenqi Li, and et~al.
\newblock {{GradViT}}: {{Gradient Inversion}} of {{Vision Transformers}}.
\newblock In {\em {{CVPR}}}, 2022.

\bibitem{hayesLOGANMembershipInference2019}
Jamie Hayes, Luca Melis, George Danezis, and Emiliano~De Cristofaro.
\newblock {{LOGAN}}: {{Membership Inference Attacks Against Generative
  Models}}.
\newblock {\em {{PoPETs}}}, 2019.

\bibitem{heDeepResidualLearning2016}
Kaiming He, Xiangyu Zhang, Shaoqing Ren, and Jian Sun.
\newblock Deep {{Residual Learning}} for {{Image Recognition}}.
\newblock In {\em {{CVPR}}}, 2016.

\bibitem{heQuantifyingMitigatingPrivacy2021}
Xinlei He and Yang Zhang.
\newblock Quantifying and {{Mitigating Privacy Risks}} of {{Contrastive
  Learning}}.
\newblock In {\em {{CCS}}}, 2021.

\bibitem{heSegmentationsLeakMembershipInference2020}
Yang He, Shadi Rahimian, Bernt Schiele, and Mario Fritz.
\newblock Segmentations-{{Leak}}: {{Membership Inference Attacks}} and
  {{Defenses}} in {{Semantic Image Segmentation}}.
\newblock In {\em {{ECCV}}}, 2020.

\bibitem{hendrycksGaussianErrorLinear2016}
Dan Hendrycks and Kevin Gimpel.
\newblock Gaussian {{Error Linear Units}} ({{GELUs}}).
\newblock {\em arXiv preprint arXiv:1606.08415}, 2016.

\bibitem{howardSearchingMobileNetV32019}
Andrew Howard, Mark Sandler, Bo~Chen, Weijun Wang, Liang-Chieh Chen, and et~al.
\newblock Searching for {{MobileNetV3}}.
\newblock In {\em {{ICCV}}}, 2019.

\bibitem{huMembershipInferenceAttacks2022}
Hongsheng Hu, Zoran Salcic, Lichao Sun, Gillian Dobbie, Philip~S. Yu, and Xuyun
  Zhang.
\newblock Membership {{Inference Attacks}} on {{Machine Learning}}: {{A
  Survey}}.
\newblock {\em ACM Comput. Surv.}, 2022.

\bibitem{huangDeepNetworksStochastic2016}
Gao Huang, Yu~Sun, Zhuang Liu, Daniel Sedra, and Kilian~Q. Weinberger.
\newblock Deep {{Networks}} with {{Stochastic Depth}}.
\newblock In {\em {{ECCV}}}, 2016.

\bibitem{jayaramanAreAttributeInference2022}
Bargav Jayaraman and David Evans.
\newblock Are {{Attribute Inference Attacks Just Imputation}}?
\newblock In {\em {{CCS}}}, pages 1569--1582, 2022.

\bibitem{jayaramanRevisitingMembershipInference2021}
Bargav Jayaraman, Lingxiao Wang, Katherine Knipmeyer, Quanquan Gu, and David
  Evans.
\newblock Revisiting {{Membership Inference Under Realistic Assumptions}}.
\newblock {\em {{PoPETs}}}, 2021.

\bibitem{kayaWhenDoesData2021}
Yigitcan Kaya and Tudor Dumitras.
\newblock When {{Does Data Augmentation Help With Membership Inference
  Attacks}}?
\newblock In {\em {{ICML}}}, 2021.

\bibitem{krizhevskyLearningMultipleLayers2009}
A.~Krizhevsky.
\newblock Learning multiple layers of features from tiny images.
\newblock 2009.

\bibitem{krizhevskyImageNetClassificationDeep2012}
Alex Krizhevsky, Ilya Sutskever, and Geoffrey~E Hinton.
\newblock {{ImageNet Classification}} with {{Deep Convolutional Neural
  Networks}}.
\newblock In {\em {{NeurIPS}}}, 2012.

\bibitem{lecunBackpropagationAppliedHandwritten1989}
Y.~LeCun, B.~Boser, J.~S. Denker, D.~Henderson, R.~E. Howard, and et~al.
\newblock Backpropagation {{Applied}} to {{Handwritten Zip Code Recognition}}.
\newblock {\em Neural Computation}, 1989.

\bibitem{leeGELUActivationFunction2023}
Minhyeok Lee.
\newblock {{GELU Activation Function}} in {{Deep Learning}}: {{A Comprehensive
  Mathematical Analysis}} and {{Performance}}.
\newblock {\em arXiv preprint arXiv:2305.12073}, 2023.

\bibitem{liMViTv2ImprovedMultiscale2022}
Yanghao Li, Chao-Yuan Wu, Haoqi Fan, Karttikeya Mangalam, and et~al.
\newblock Mvitv2: Improved multiscale vision transformers for classification
  and detection.
\newblock In {\em {{CVPR}}}, 2022.

\bibitem{liMembershipLeakageLabelOnly2021}
Zheng Li and Yang Zhang.
\newblock Membership {{Leakage}} in {{Label-Only Exposures}}.
\newblock In {\em {{CCS}}}, 2021.

\bibitem{liAuditingPrivacyDefenses2022}
Zhuohang Li, Jiaxin Zhang, Luyang Liu, and Jian Liu.
\newblock Auditing {{Privacy Defenses}} in {{Federated Learning}} via
  {{Generative Gradient Leakage}}.
\newblock In {\em {{CVPR}}}, 2022.

\bibitem{liuWhenMachineLearning2021}
Bo~Liu, Ming Ding, Sina Shaham, Wenny Rahayu, Farhad Farokhi, and Zihuai Lin.
\newblock When {{Machine Learning Meets Privacy}}: {{A Survey}} and
  {{Outlook}}.
\newblock {\em ACM Comput. Surv.}, 2021.

\bibitem{liuMLDoctorHolisticRisk2022}
Yugeng Liu, Rui Wen, Xinlei He, Ahmed Salem, Zhikun Zhang, and et~al.
\newblock {{ML-Doctor}}: {{Holistic Risk Assessment}} of {{Inference Attacks
  Against Machine Learning Models}}.
\newblock In {\em {{USENIX Security}}}, 2022.

\bibitem{liuSwinTransformerHierarchical2021}
Ze~Liu, Yutong Lin, Yue Cao, Han Hu, Yixuan Wei, and et~al.
\newblock Swin {{Transformer}}: {{Hierarchical Vision Transformer}} using
  {{Shifted Windows}}.
\newblock In {\em {{ICCV}}}, 2021.

\bibitem{liuConvNet2020s2022}
Zhuang Liu, Hanzi Mao, Chao-Yuan Wu, Christoph Feichtenhofer, and et~al.
\newblock A {{ConvNet}} for the 2020s.
\newblock In {\em {{CVPR}}}, 2022.

\bibitem{liuDeepLearningFace2015}
Ziwei Liu, Ping Luo, Xiaogang Wang, and Xiaoou Tang.
\newblock Deep {{Learning Face Attributes}} in the {{Wild}}.
\newblock In {\em {{ICCV}}}, 2015.

\bibitem{longPragmaticApproachMembership2020}
Yunhui Long, Lei Wang, Diyue Bu, Vincent Bindschaedler, Xiaofeng Wang, and
  et~al.
\newblock A {{Pragmatic Approach}} to {{Membership Inferences}} on {{Machine
  Learning Models}}.
\newblock In {\em {{EuroS}}\&{{P}}}, 2020.

\bibitem{luAPRILFindingAchilles2022}
Jiahao Lu, Xi~Sheryl Zhang, Tianli Zhao, Xiangyu He, and Jian Cheng.
\newblock {{APRIL}}: {{Finding}} the {{Achilles}}' {{Heel}} on {{Privacy}} for
  {{Vision Transformers}}.
\newblock In {\em {{CVPR}}}, 2022.

\bibitem{luoUnderstandingEffectiveReceptive2016}
Wenjie Luo, Yujia Li, Raquel Urtasun, and Richard Zemel.
\newblock Understanding the {{Effective Receptive Field}} in {{Deep
  Convolutional Neural Networks}}.
\newblock In {\em {{NeurIPS}}}, 2016.

\bibitem{mehnazAreYourSensitive2022}
Shagufta Mehnaz, Sayanton~V. Dibbo, Ehsanul Kabir, Ninghui Li, and Elisa
  Bertino.
\newblock Are {{Your Sensitive Attributes Private}}? {{Novel Model Inversion
  Attribute Inference Attacks}} on {{Classification Models}}.
\newblock In {\em {{USENIX Security}}}, 2022.

\bibitem{melisExploitingUnintendedFeature2019}
Luca Melis, Congzheng Song, Emiliano De~Cristofaro, and Vitaly Shmatikov.
\newblock Exploiting {{Unintended Feature Leakage}} in {{Collaborative
  Learning}}.
\newblock In {\em {{S\&P}}}, 2019.

\bibitem{nasrScalableExtractionTraining2023}
Milad Nasr, Nicholas Carlini, Jonathan Hayase, Matthew Jagielski, and et~al.
\newblock Scalable {{Extraction}} of {{Training Data}} from ({{Production}})
  {{Language Models}}.
\newblock {\em arXiv preprint arXiv:2311.17035}, 2023.

\bibitem{nasrComprehensivePrivacyAnalysis2019}
Milad Nasr, Reza Shokri, and Amir Houmansadr.
\newblock Comprehensive {{Privacy Analysis}} of {{Deep Learning}}: {{Passive}}
  and {{Active White-box Inference Attacks}} against {{Centralized}} and
  {{Federated Learning}}.
\newblock In {\em {{S\&P}}}, 2019.

\bibitem{paulVisionTransformersAre2022}
Sayak Paul and Pin-Yu Chen.
\newblock Vision {{Transformers Are Robust Learners}}.
\newblock {\em {{AAAI}}}, 2022.

\bibitem{radosavovicDesigningNetworkDesign2020}
Ilija Radosavovic, Raj~Prateek Kosaraju, Ross Girshick, Kaiming He, and Piotr
  Dollar.
\newblock Designing {{Network Design Spaces}}.
\newblock In {\em {{CVPR}}}, 2020.

\bibitem{raghuVisionTransformersSee2021}
Maithra Raghu, Thomas Unterthiner, Simon Kornblith, Chiyuan Zhang, and Alexey
  Dosovitskiy.
\newblock Do {{Vision Transformers See Like Convolutional Neural Networks}}?
\newblock In {\em {{NeurIPS}}}, 2021.

\bibitem{rajputExponentialImprovementMemorization2021}
Shashank Rajput, Kartik Sreenivasan, Dimitris Papailiopoulos, and Amin Karbasi.
\newblock An {{Exponential Improvement}} on the {{Memorization Capacity}} of
  {{Deep Threshold Networks}}.
\newblock In {\em {{NeurIPS}}}, 2021.

\bibitem{sablayrollesWhiteboxVsBlackbox2019}
Alexandre Sablayrolles, Matthijs Douze, Cordelia Schmid, Yann Ollivier, and
  Herve Jegou.
\newblock White-box vs {{Black-box}}: {{Bayes Optimal Strategies}} for
  {{Membership Inference}}.
\newblock In {\em {{ICML}}}, 2019.

\bibitem{salemMLLeaksModelData2019}
Ahmed Salem, Yang Zhang, Mathias Humbert, Pascal Berrang, Mario Fritz, and
  Michael Backes.
\newblock {{ML-Leaks}}: {{Model}} and data independent membership inference
  attacks and defenses on machine learning models.
\newblock In {\em {{NDSS}}}, 2019.

\bibitem{shejwalkarMembershipPrivacyMachine2021}
Virat Shejwalkar and Amir Houmansadr.
\newblock Membership {{Privacy}} for {{Machine Learning Models Through
  Knowledge Transfer}}.
\newblock {\em AAAI}, 2021.

\bibitem{shokriMembershipInferenceAttacks2017}
Reza Shokri, Marco Stronati, Congzheng Song, and Vitaly Shmatikov.
\newblock Membership {{Inference Attacks Against Machine Learning Models}}.
\newblock In {\em {{S\&P}}}, 2017.

\bibitem{simonyanVeryDeepConvolutional2015}
Karen Simonyan and Andrew Zisserman.
\newblock Very {{Deep Convolutional Networks}} for {{Large-Scale Image
  Recognition}}.
\newblock {\em arXiv preprint arXiv:1409.1556}, 2015.

\bibitem{songInformationLeakageEmbedding2020}
Congzheng Song and Ananth Raghunathan.
\newblock Information {{Leakage}} in {{Embedding Models}}.
\newblock In {\em {{CCS}}}, 2020.

\bibitem{songOverlearningRevealsSensitive2020}
Congzheng Song and Vitaly Shmatikov.
\newblock Overlearning {{Reveals Sensitive Attributes}}.
\newblock In {\em {{ICLR}}}, 2020.

\bibitem{songSystematicEvaluationPrivacy2021}
Liwei Song and Prateek Mittal.
\newblock Systematic {{Evaluation}} of {{Privacy Risks}} of {{Machine Learning
  Models}}.
\newblock In {\em {{USENIX Security}}}, 2021.

\bibitem{steinerHowTrainYour2022}
Andreas~Peter Steiner, Alexander Kolesnikov, Xiaohua Zhai, Ross Wightman, and
  et~al.
\newblock How to train your {{ViT}}? {{Data}}, augmentation, and regularization
  in vision transformers.
\newblock {\em {{TMLR}}}, 2022.

\bibitem{szegedyGoingDeeperConvolutions2015}
Christian Szegedy, Wei Liu, Yangqing Jia, Pierre Sermanet, Scott Reed, and
  et~al.
\newblock Going deeper with convolutions.
\newblock In {\em {{CVPR}}}, 2015.

\bibitem{szegedyRethinkingInceptionArchitecture2016}
Christian Szegedy, Vincent Vanhoucke, Sergey Ioffe, Jon Shlens, and Zbigniew
  Wojna.
\newblock Rethinking the {{Inception Architecture}} for {{Computer Vision}}.
\newblock In {\em {{CVPR}}}, 2016.

\bibitem{tanEfficientNet2019}
Mingxing Tan and Quoc Le.
\newblock {E}fficient{N}et: Rethinking model scaling for convolutional neural
  networks.
\newblock In {\em {{ICML}}}, 2019.

\bibitem{touvronTrainingDataefficientImage2021}
Hugo Touvron, Matthieu Cord, Matthijs Douze, Francisco Massa, Alexandre
  Sablayrolles, and Herve Jegou.
\newblock Training data-efficient image transformers \& distillation through
  attention.
\newblock In {\em {{ICML}}}, 2021.

\bibitem{touvronGoingDeeperImage2021}
Hugo Touvron, Matthieu Cord, Alexandre Sablayrolles, Gabriel Synnaeve, and
  Herv{\'e} J{\'e}gou.
\newblock Going deeper with {{Image Transformers}}.
\newblock In {\em {{ICCV}}}, 2021.

\bibitem{truexDemystifyingMembershipInference2019}
Stacey Truex, Ling Liu, Mehmet~Emre Gursoy, Lei Yu, and Wenqi Wei.
\newblock Demystifying {{Membership Inference Attacks}} in {{Machine Learning}}
  as a {{Service}}.
\newblock {\em IEEE Transactions on Services Computing}, 2019.

\bibitem{vaswaniAttentionAllYou2017}
Ashish Vaswani, Noam Shazeer, Niki Parmar, Jakob Uszkoreit, Llion Jones, and
  et~al.
\newblock Attention is {{All}} you {{Need}}.
\newblock In {\em {{NeurIPS}}}, 2017.

\bibitem{wangCanCNNsBe2023}
Zeyu Wang, Yutong Bai, Yuyin Zhou, and Cihang Xie.
\newblock Can {{CNNs Be More Robust Than Transformers}}?
\newblock In {\em {{ICLR}}}, 2023.

\bibitem{wangImageQualityAssessment2004}
Zhou Wang, A.C. Bovik, H.R. Sheikh, and E.P. Simoncelli.
\newblock Image quality assessment: From error visibility to structural
  similarity.
\newblock {\em IEEE Transactions on Image Processing}, 2004.

\bibitem{watsonImportanceDifficultyCalibration2022}
Lauren Watson, Chuan Guo, Graham Cormode, and Alexandre Sablayrolles.
\newblock On the {{Importance}} of {{Difficulty Calibration}} in {{Membership
  Inference Attacks}}.
\newblock In {\em {{ICLR}}}, 2022.

\bibitem{xiaoEarlyConvolutionsHelp2021}
Tete Xiao, Mannat Singh, Eric Mintun, Trevor Darrell, Piotr Dollar, and Ross
  Girshick.
\newblock Early {{Convolutions Help Transformers See Better}}.
\newblock In {\em {{NeurIPS}}}, 2021.

\bibitem{xieAggregatedResidualTransformations2017}
Saining Xie, Ross Girshick, Piotr Doll{\'a}r, Zhuowen Tu, and Kaiming He.
\newblock Aggregated {{Residual Transformations}} for {{Deep Neural Networks}}.
\newblock In {\em {{CVPR}}}, 2017.

\bibitem{xuUnderstandingImprovingLayer2019}
Jingjing Xu, Xu~Sun, Zhiyuan Zhang, Guangxiang Zhao, and et~al.
\newblock Understanding and {{Improving Layer Normalization}}.
\newblock In {\em {{NeurIPS}}}, 2019.

\bibitem{yangXLNetGeneralizedAutoregressive2019}
Zhilin Yang, Zihang Dai, Yiming Yang, Jaime Carbonell, Russ~R Salakhutdinov,
  and Quoc~V Le.
\newblock {{XLNet}}: {{Generalized Autoregressive Pretraining}} for {{Language
  Understanding}}.
\newblock In {\em {{NeurIPS}}}, 2019.

\bibitem{yeEnhancedMembershipInference2022}
Jiayuan Ye, Aadyaa Maddi, Sasi~Kumar Murakonda, Vincent Bindschaedler, and Reza
  Shokri.
\newblock Enhanced {{Membership Inference Attacks}} against {{Machine Learning
  Models}}.
\newblock In {\em {{CCS}}}, 2022.

\bibitem{yeomPrivacyRiskMachine2018}
Samuel Yeom, Irene Giacomelli, Matt Fredrikson, and Somesh Jha.
\newblock Privacy {{Risk}} in {{Machine Learning}}: {{Analyzing}} the
  {{Connection}} to {{Overfitting}}.
\newblock In {\em {{CSF}}}, 2018.

\bibitem{yinSeeGradientsImage2021}
Hongxu Yin, Arun Mallya, Arash Vahdat, Jose~M. Alvarez, Jan Kautz, and Pavlo
  Molchanov.
\newblock See {{Through Gradients}}: {{Image Batch Recovery}} via
  {{GradInversion}}.
\newblock In {\em {{CVPR}}}, 2021.

\bibitem{yuanTokenstoTokenViTTraining2021}
Li~Yuan, Yunpeng Chen, Tao Wang, Weihao Yu, Yujun Shi, and et~al.
\newblock Tokens-to-{{Token ViT}}: {{Training Vision Transformers}} from
  {{Scratch}} on {{ImageNet}}.
\newblock In {\em {{ICCV}}}, 2021.

\bibitem{zhangHowDoesDeep2024}
Guangsheng Zhang, Bo~Liu, Huan Tian, Tianqing Zhu, Ming Ding, and Wanlei Zhou.
\newblock How {{Does}} a {{Deep Learning Model Architecture Impact Its
  Privacy}}? {{A Comprehensive Study}} of {{Privacy Attacks}} on {{CNNs}} and
  {{Transformers}}.
\newblock {\em arXiv preprint arXiv:2210.11049}, 2024.

\bibitem{zhangLabelOnlyMembershipInference2022}
Guangsheng Zhang, Bo~Liu, Tianqing Zhu, Ming Ding, and Wanlei Zhou.
\newblock Label-{{Only Membership Inference Attacks}} and {{Defenses In
  Semantic Segmentation Models}}.
\newblock {\em IEEE Transactions on Dependable and Secure Computing}, 2022.

\bibitem{zhangVisualPrivacyAttacks2022}
Guangsheng Zhang, Bo~Liu, Tianqing Zhu, Andi Zhou, and Wanlei Zhou.
\newblock Visual privacy attacks and defenses in deep learning: A survey.
\newblock {\em Artif Intell Rev}, 2022.

\bibitem{zhangUnreasonableEffectivenessDeep2018}
Richard Zhang, Phillip Isola, Alexei~A. Efros, Eli Shechtman, and Oliver Wang.
\newblock The {{Unreasonable Effectiveness}} of {{Deep Features}} as a
  {{Perceptual Metric}}.
\newblock In {\em {{CVPR}}}, 2018.

\bibitem{zhangSurveyGradientInversion2022}
Rui Zhang, Song Guo, Junxiao Wang, Xin Xie, and Dacheng Tao.
\newblock A {{Survey}} on {{Gradient Inversion}}: {{Attacks}}, {{Defenses}} and
  {{Future Directions}}.
\newblock In {\em {{IJCAI}}}, 2022.

\bibitem{zhaoFeasibilityAttributeInference2021}
Benjamin Zi~Hao Zhao, Aviral Agrawal, Catisha Coburn, Hassan~Jameel Asghar, and
  et~al.
\newblock On the ({{In}}){{Feasibility}} of {{Attribute Inference Attacks}} on
  {{Machine Learning Models}}.
\newblock In {\em {{EuroS}}\&{{P}}}, 2021.

\bibitem{zhaoIDLGImprovedDeep2020}
Bo~Zhao, Konda~Reddy Mopuri, and Hakan Bilen.
\newblock {{iDLG}}: {{Improved Deep Leakage}} from {{Gradients}}.
\newblock {\em arXiv preprint arXiv:2001.02610}, 2020.

\bibitem{zhuDeepLeakageGradients2019}
Ligeng Zhu, Zhijian Liu, and Song Han.
\newblock Deep {{Leakage}} from {{Gradients}}.
\newblock {\em {{NeurIPS}}}, 2019.

\bibitem{zouPrivacyAnalysisDeep2020}
Yang Zou, Zhikun Zhang, Michael Backes, and Yang Zhang.
\newblock Privacy {{Analysis}} of {{Deep Learning}} in the {{Wild}}:
  {{Membership Inference Attacks}} against {{Transfer Learning}}.
\newblock {\em arXiv preprint arXiv:2009.04872}, 2020.

\end{thebibliography}
}

\appendix

\section{Detailed Architecture Specifications for Experiments}
\label{sec:detailed_arch}

\begin{table*}[t]
\scriptsize
  \centering
  \caption{Detailed Architecture Specifications for 14 steps in~\Cref{sec:impact_of_micro}.}
  \resizebox{\linewidth}{!}{
    \begin{tabular}{c|c|c|c|c|c|c|c}
    \toprule
          & ResNet-50, Step 1 & Step 2 & Step 3 & Step 4 & Step 5 & Step 6 & Step 7 \\
    \midrule
    \multirow{2}{*}{stem}   & 7 × 7, 64, stride 2    & 7 × 7, \zb{96}, stride 2    & 7 × 7, 96, stride 2    & \zb{4 × 4}, 96, stride \zb{4}    & 4 × 4, 96, stride 4    & 4 × 4, 96, stride 4    & 4 × 4, 96, stride 4 \\
                            & 3 × 3, maxpl, stride 2 & 3 × 3, maxpl, stride 2      & 3 × 3, maxpl, stride 2 & 3 × 3, maxpl, stride 2           & 3 × 3, maxpl, stride 2 & 3 × 3, maxpl, stride 2 & 3 × 3, maxpl, stride 2 \\
    \midrule
    \multirow{3}{*}{block1} & [1 × 1, 64       & [1 × 1, \zb{96}       & [1 × 1, 96            & [1 × 1, 96       & [1 × 1, 96       & [1 × 1, 96           & [\zb{d7 × 7}, 96 \\
                            & 3 × 3, 64        & 3 × 3, \zb{96}        & 3 × 3, 96             & 3 × 3, 96        & \zb{d3 × 3}, 96  & d3 × 3, \zb{384}     & 1 × 1, 384 \\
                            & 1 × 1, 256] × 3  & 1 × 1, \zb{384}] × 3  & 1 × 1, 384] × \zb{3}  & 1 × 1, 384] × 3  & 1 × 1, 384] × 3  & 1 × 1, \zb{96}] × 3  & 1 × 1, 96] × 3 \\
    \midrule
    \multirow{3}{*}{block2} & [1 × 1, 128      & [1 × 1, \zb{192}      & [1 × 1, 192           & [1 × 1, 192      & [1 × 1, 192      & [1 × 1, 192          & [\zb{d7 × 7}, 192 \\
                            & 3 × 3, 128       & 3 × 3, \zb{192}       & 3 × 3, 192            & 3 × 3, 192       & \zb{d3 × 3}, 192 & d3 × 3, \zb{768}     & 1 × 1, 768 \\
                            & 1 × 1, 512] × 4  & 1 × 1, \zb{768}] × 4  & 1 × 1, 768] × \zb{3}  & 1 × 1, 768] × 3  & 1 × 1, 768] × 3  & 1 × 1, \zb{192}] × 3 & 1 × 1, 192] × 3 \\
    \midrule
    \multirow{3}{*}{block3} & [1 × 1, 256      & [1 × 1, \zb{384}      & [1 × 1, 384           & [1 × 1, 384      & [1 × 1, 384      & [1 × 1, 384          & [\zb{d7 × 7}, 384 \\
                            & 3 × 3, 256       & 3 × 3, \zb{384}       & 3 × 3, 384            & 3 × 3, 384       & \zb{d3 × 3}, 384 & d3 × 3, \zb{1536}    & 1 × 1, 1536 \\
                            & 1 × 1, 1024] × 6 & 1 × 1, \zb{1536}] × 6 & 1 × 1, 1536] × \zb{9} & 1 × 1, 1536] × 9 & 1 × 1, 1536] × 9 & 1 × 1, \zb{384}] × 9 & 1 × 1, 384] × 9 \\
    \midrule
    \multirow{3}{*}{block4} & [1 × 1, 512      & [1 × 1, \zb{768}      & [1 × 1, 768           & [1 × 1, 768      & [1 × 1, 768      & [1 × 1, 768          & [\zb{d7 × 7}, 768 \\
                            & 3 × 3, 512       & 3 × 3, \zb{768}       & 3 × 3, 768            & 3 × 3, 768       & \zb{d3 × 3}, 768 & d3 × 3, \zb{3072}    & 1 × 1, 3072 \\
                            & 1 × 1, 2048] × 3 & 1 × 1, \zb{3072}] × 3 & 1 × 1, 3072] × \zb{3} & 1 × 1, 3072] × 3 & 1 × 1, 3072] × 3 & 1 × 1, \zb{768}] × 3 & 1 × 1, 768] × 3 \\
    \midrule
    other specs             & ReLU, BN & ReLU, BN & ReLU, BN & ReLU, BN & ReLU, BN & ReLU, BN & ReLU, BN \\
    \midrule
    - & - & - & - & - & - & - & - \\
    \midrule
                            & Step8 & Step9 & Step10 & Step11 & Step12 & Step13 & ConvNeXt-T, Step14 \\
    \midrule
    \multirow{2}{*}{stem}   & 4 × 4, 96, stride 4 & 4 × 4, 96, stride 4 & 4 × 4, 96, stride 4 & 4 × 4, 96, stride 4 & 4 × 4, 96, stride 4 & 4 × 4, 96, stride 4 & 4 × 4, 96, stride 4 \\
                            & \zb{(removed)} &       &       &       &       &       &  \\
    \midrule
    \multirow{3}{*}{block1} & [d7 × 7, 96     & [d7 × 7, 96     & [d7 × 7, 96     & [d7 × 7, 96     & [d7 × 7, 96     & [d7 × 7, 96     & [d7 × 7, 96 \\
                            & 1 × 1, 384      & 1 × 1, 384      & 1 × 1, 384      & 1 × 1, 384      & 1 × 1, 384      & 1 × 1, 384      & 1 × 1, 384 \\
                            & 1 × 1, 96] × 3  & 1 × 1, 96] × 3  & 1 × 1, 96] × 3  & 1 × 1, 96] × 3  & 1 × 1, 96] × 3  & 1 × 1, 96] × 3  & 1 × 1, 96] × 3 \\
    \midrule
    sep ds                  &       &       &       &       &       & \zb{2 × 2, 192, stride 2} & 2 × 2, 192, stride 2 \\
    \midrule
    \multirow{3}{*}{block2} & [d7 × 7, 192    & [d7 × 7, 192    & [d7 × 7, 192    & [d7 × 7, 192    & [d7 × 7, 192    & [d7 × 7, 192    & [d7 × 7, 192 \\
                            & 1 × 1, 768      & 1 × 1, 768      & 1 × 1, 768      & 1 × 1, 768      & 1 × 1, 768      & 1 × 1, 768      & 1 × 1, 768 \\
                            & 1 × 1, 192] × 3 & 1 × 1, 192] × 3 & 1 × 1, 192] × 3 & 1 × 1, 192] × 3 & 1 × 1, 192] × 3 & 1 × 1, 192] × 3 & 1 × 1, 192] × 3 \\
    \midrule
    sep ds                  &       &       &       &       &       & \zb{2 × 2, 192, stride 2} & 2 × 2, 192, stride 2 \\
    \midrule
    \multirow{3}{*}{block3} & [d7 × 7, 384    & [d7 × 7, 384    & [d7 × 7, 384    & [d7 × 7, 384    & [d7 × 7, 384    & [d7 × 7, 384    & [d7 × 7, 384 \\
                            & 1 × 1, 1536     & 1 × 1, 1536     & 1 × 1, 1536     & 1 × 1, 1536     & 1 × 1, 1536     & 1 × 1, 1536     & 1 × 1, 1536 \\
                            & 1 × 1, 384] × 9 & 1 × 1, 384] × 9 & 1 × 1, 384] × 9 & 1 × 1, 384] × 9 & 1 × 1, 384] × 9 & 1 × 1, 384] × 9 & 1 × 1, 384] × 9 \\
    \midrule
    sep ds                  &       &       &       &       &       & \zb{2 × 2, 192, stride 2} & 2 × 2, 192, stride 2 \\
    \midrule
    \multirow{3}{*}{block4} & [d7 × 7, 768    & [d7 × 7, 768    & [d7 × 7, 768    & [d7 × 7, 768    & [d7 × 7, 768    & [d7 × 7, 768    & [d7 × 7, 768 \\
                            & 1 × 1, 3072     & 1 × 1, 3072     & 1 × 1, 3072     & 1 × 1, 3072     & 1 × 1, 3072     & 1 × 1, 3072     & 1 × 1, 3072 \\
                            & 1 × 1, 768] × 3 & 1 × 1, 768] × 3 & 1 × 1, 768] × 3 & 1 × 1, 768] × 3 & 1 × 1, 768] × 3 & 1 × 1, 768] × 3 & 1 × 1, 768] × 3 \\
    \midrule
    \multirow{4}{*}{other specs} & ReLU, BN & \zb{GELU}, BN & \zb{Fewer GELU}, BN & Fewer GELU    & Fewer GELU             & Fewer GELU,       & Fewer GELU  \\
                &          &               &                     & \zb{Fewer BN} & \zb{Fewer LN}          & Fewer LN          & Fewer LN \\
                &          &               &                     &               & \zb{Conv w/ True bias} & Conv w/ True bias & Conv w/ True bias \\
                &          &               &                     &               &                        &                   & \zb{StoDepth, LayerScale} \\
    \bottomrule
    \end{tabular}}%
  \label{tab:detailed_architectures}%
\end{table*}%

We present a detailed architecture comparison between models from 14 steps (changing from ResNet-50 to ConvNeXt-T) in~\Cref{tab:detailed_architectures}. Changes are marked in bold.

\begin{table}[h]
\scriptsize
\centering
  \caption{The results of membership inference attacks on more models of CNNs and Transformers on CIFAR10 as a supplement for~\Cref{sec:exp_mia}.}
    \begin{tabular}{c|c|c}
    \toprule
    & Task acc $\uparrow$ & Attack acc $\uparrow$ \\
    \midrule
    ResNet-50 & 0.8220 $\pm$ 0.0023 & 0.6385 $\pm$ 0.0078 \\
    EfficientNet-B4 & 0.8180 $\pm$ 0.0052 & 0.6115 $\pm$ 0.0049 \\
    RegNetX-4.0GF & 0.8326 $\pm$ 0.0018 & 0.6197 $\pm$ 0.0042 \\
    \midrule
    Swin-T & 0.8335 $\pm$ 0.0042 & 0.6904 $\pm$ 0.0052 \\
    MViTV2-T & 0.8279 $\pm$ 0.0057 & 0.6813 $\pm$ 0.0056 \\
    DaViT-T & 0.8311 $\pm$ 0.0045 & 0.6752 $\pm$ 0.0039 \\
    \bottomrule
    \end{tabular}%
  \label{tab:mia_comp_results_more_models}%
\end{table}%

\begin{table}[h]
\scriptsize
\centering
  \caption{The results of attribute inference attacks on more models of CNNs and Transformers on CelebA as a supplement for~\Cref{sec:exp_aia}.}
    \begin{tabular}{c|c|cc}
    \toprule
    & Task acc $\uparrow$ & Attack acc $\uparrow$ & Macro F1 $\uparrow$ \\
    \midrule
    ResNet-50 & 0.6666 $\pm$ 0.0020 & 0.6854 $\pm$ 0.0015 & 0.3753 $\pm$ 0.0012 \\
    EfficientNet-B4 & 0.6192 $\pm$ 0.0038 & 0.6504 $\pm$ 0.0031 & 0.4036 $\pm$ 0.0018 \\
    RegNetX-4.0GF & 0.6248 $\pm$ 0.0041 & 0.6701 $\pm$ 0.0016 & 0.3649 $\pm$ 0.0022 \\
    \midrule
    Swin-T & 0.6587 $\pm$ 0.0023 & 0.7312 $\pm$ 0.0014 & 0.5530 $\pm$ 0.0019 \\
    MViTV2-T & 0.6151 $\pm$ 0.0012 & 0.6908 $\pm$ 0.0025 & 0.4883 $\pm$ 0.0024 \\
    DaViT-T & 0.6517 $\pm$ 0.0030 & 0.7263 $\pm$ 0.0015 & 0.5093 $\pm$ 0.0011 \\
    \bottomrule
    \end{tabular}%
  \label{tab:aia_comp_results_more_models}%
\end{table}%

\begin{table}[h]
  \scriptsize
    \centering
    \caption{The results of gradient inversion attacks on more models of CNNs and Transformers on CIFAR10 as a supplement for~\Cref{sec:exp_gradient_inversion}.}
    \resizebox{\linewidth}{!}{
    \begin{tabular}{c|cccc}
    \toprule
         &  MSE $\downarrow$ &  PSNR $\uparrow$ &  LPIPS $\downarrow$ & SSIM $\uparrow$ \\
    \midrule
         ResNet-50 & 1.3308 $\pm$ 0.6507 & 11.30 $\pm$ 2.24 & 0.1143 $\pm$ 0.0403 & 0.0946 $\pm$ 0.0989 \\
         EfficientNet-B4 & 0.8178 $\pm$ 0.7276 & 14.6970 $\pm$ 3.9812 & 0.1953 $\pm$ 0.0816 & 0.2698 $\pm$ 0.1413\\
         RegNetX-4.0GF & 0.7937 $\pm$ 0.4853 & 13.8190 $\pm$ 2.6519 & 0.0851 $\pm$ 0.0391 & 0.3197 $\pm$ 0.1282 \\
    \midrule
         Swin-T & 0.0069 $\pm$ 0.0071 & 36.24 $\pm$ 5.21 & 0.0012 $\pm$ 0.0016 & 0.9892 $\pm$ 0.0118 \\
         MViTV2-T & 0.0711 $\pm$ 0.0172 & 23.6450 $\pm$ 1.1224 & 0.0068 $\pm$ 0.0033 & 0.8639 $\pm$ 0.0463 \\
         DaViT-T & 0.0615 $\pm$ 0.0425 & 25.9200 $\pm$ 4.7786 & 0.0090 $\pm$ 0.0076 & 0.9347 $\pm$ 0.0489 \\
    \bottomrule
    \end{tabular}}
    \label{tab:inv_comp_results_more_models}
\end{table}

\begin{figure}[h]
  \begin{subfigure}[t]{.23\textwidth}
    \centering
    \includegraphics[width=\linewidth]{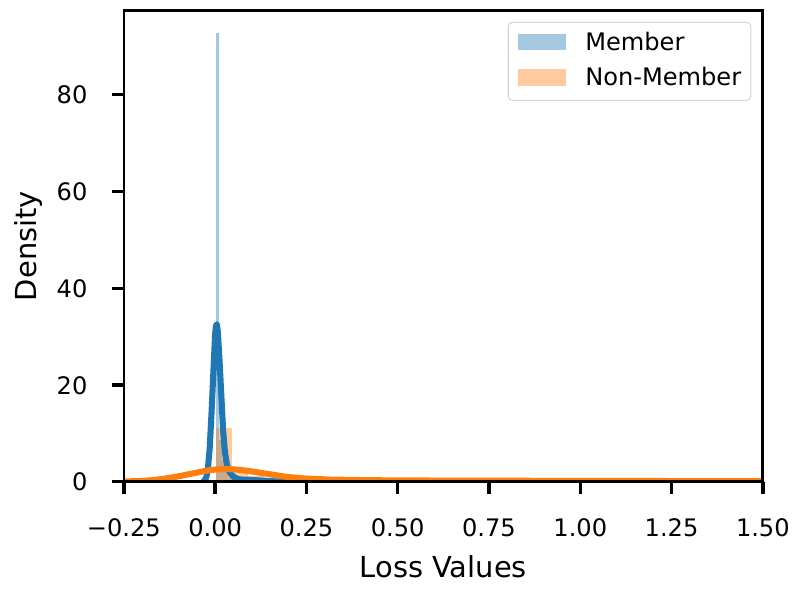}
    \caption{Loss distributions on ResNet-50}
    \label{fig:loss_hist_res50}
  \end{subfigure}
  \hfill
  \begin{subfigure}[t]{.23\textwidth}
    \centering
    \includegraphics[width=\linewidth]{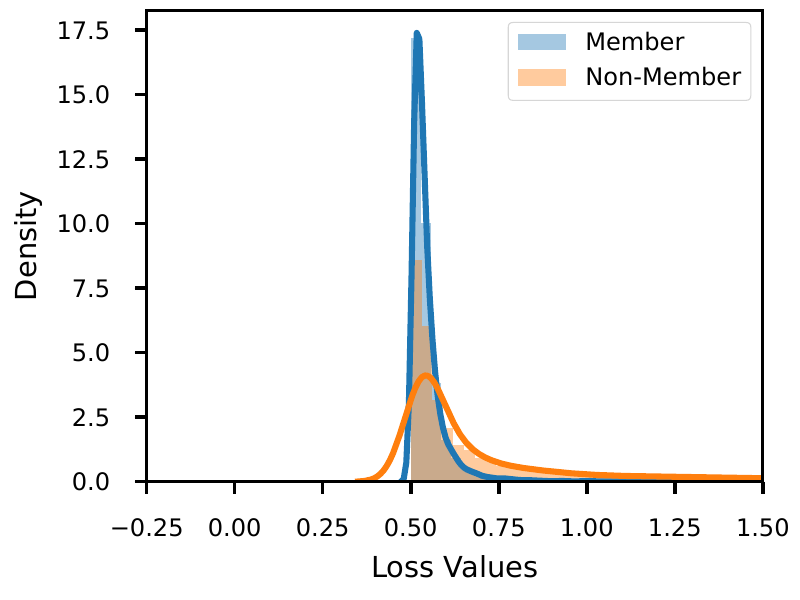}
    \caption{Loss distributions on Swin-T}
    \label{fig:loss_hist_swint}
  \end{subfigure}
  \caption{The loss distributions of membership inference attacks against ResNet-50 and Swin-T on CIFAR10.}
  \label{fig:loss_hist}
\end{figure}

\begin{table}[t]
  \scriptsize
  \caption{The performance of gradient inversion attacks when segmenting Swin-T to make a selection of gradients as a supplement for~\Cref{sec:segmenting_trans}.}
  \label{tab:gradient_sel_swint}
  \centering
  \begin{tabular}{lcccc}
    \toprule
    Layers & Num of layers & Params & MSE $\downarrow$ & PSNR $\uparrow$ \\
    \midrule
    All       & 173 & 27.51M & 0.0069 $\pm$ 0.0071 & 36.24 $\pm$ 5.21 \\
    Stem      & 4   & 0.01M & 0.0073 $\pm$ 0.0058 & 34.38 $\pm$ 2.89 \\
    Attention & 48  & 8.64M & 0.2691 $\pm$ 0.2099 & 17.72 $\pm$ 1.21 \\
    MLP       & 68  & 18.82M & 0.3132 $\pm$ 0.1001 & 17.06 $\pm$ 3.03 \\
    Norm      & 49  & 0.03M & 0.3526 $\pm$ 0.0826 & 16.65 $\pm$ 0.95 \\
    Head      & 4   & 0.01M & 1.1179 $\pm$ 0.2996 & 11.71 $\pm$ 1.29 \\
    \bottomrule
  \end{tabular}
\end{table}

\section{Additional Experimental Results}

\Cref{tab:mia_comp_results_more_models,tab:aia_comp_results_more_models,tab:inv_comp_results_more_models} show privacy attack results for more models of CNNs and Transformers. 
The comparison still involves models with similar parameter sizes, and our conclusion remains unchanged that Transformers exhibit higher attack performance.

We plot the loss distributions between the member and non-member data in membership inference attacks for both CNNs and Transformers in~\Cref{fig:loss_hist}.
We can see that the distributions of member and non-member data for CNNs are more concentrated at 0,
and the distributions for Transformers are concentrated at 0.5.

\Cref{tab:gradient_sel_swint} demonstrate the additional results of gradient inversion attacks when segmenting Swin-T to make a selection of gradients.

\section{More Theoretical Discussion}
\label{sec:more_theoretical_discussion}

In~\Cref{sec:discussion}, 
we offer general discussions on the privacy impact of four components in Transformers. 
In this section, 
we delve deeper into more theoretical discussions.

\subsection{The Impact of Attention Modules}

In~\Cref{sec:segmenting_trans,sec:discussion_attention}, 
we show that the utilization of attention modules makes Transformers more susceptible to privacy leakage than CNNs.
This susceptibility is attributed to the larger receptive field of attention modules, contributing to increased model memorization and, subsequently, heightened privacy leakage. 
Let us explore this theoretical framework step by step:

\paragraph{Receptive Field.}
Given a neural network with $n$ layers, the output of the $i$-th layer ($i \in \{ 1, ..., n\}$) is the feature map $f_i$.
We have the input image as $f_1$ and the final output feature map as $f_n$.
We define $\mathcal{R}_i$ as the receptive field size of feature map $f_i$.
In other words, 
$\mathcal{R}_i$ is related to the process of the $i$-th layer.
The receptive field $\mathcal{R}_i$ can be defined as
\begin{equation}
   \mathcal{R}_i = \zeta (\mathcal{R}_{i + 1}, \alpha_1, \alpha_2, ..., \alpha_n),
\label{eq:rf} 
\end{equation}
where $\alpha_1, \alpha_2, ..., \alpha_n$ are calculation factors in $\mathcal{R}_{i + 1}$, and $\zeta$ is the calculation process of $\mathcal{R}_{i + 1}$.

In a CNN model, 
the receptive field of a neuron represents the convolution-computation region in the input space that influences the output of that neuron. 
It signifies the area of the input data to which the neuron reacts.
According to~\cite{araujoComputingReceptiveFields2019}, 
the receptive field in a CNN is determined by:
\begin{equation}
    \mathcal{R}_{i}^{C} = s_{i + 1} \cdot \mathcal{R}_{i + 1}^{C} + (k_{i + 1} - s_{i + 1}),
\label{eq:rfc}
\end{equation}
where $k_i$ and $s_i$ are the kernel size and the stride in a convolution layer.
Based on~\Cref{eq:rf,eq:rfc}, 
we can define the receptive field size of a neuron in a CNN layer $\mathcal{R}^{C}$ as follows (we omit $i$ for the simplicity of notations):
\begin{equation}
    \mathcal{R}^{C} = \zeta (k, s),
\label{eq:rfc1}
\end{equation}
where $k$ and $s$ are the kernel size and the stride.

The receptive field in a Transformer model, 
utilizing attention mechanisms, 
significantly differs from CNNs. 
In Transformers, 
attention layers calculate each position in the input sequence using query ($q$), key ($k$), and value ($v$) matrices.
The output of the attention module is computed as follows:
\begin{equation}
    \text{Attention} (q, k, v) = \text{SoftMax} (q \cdot k^T / \sqrt{d}) \cdot v,
\label{eq:rft}
\end{equation}
where $d$ is the scaling factor based on query and key.
Based on~\Cref{eq:rf,eq:rft}, we can define the size of the receptive field of an attention layer $\mathcal{R}^{T}$ as:
\begin{equation}
    \mathcal{R}^{T} = \zeta (q, k, v).
\label{eq:rft1}
\end{equation}
As $q$, $k$, and $v$ in a Transformer model are derived from the entire input sequence, $\mathcal{R}^{T}$ tends to be much larger.
With~\Cref{eq:rfc1,eq:rft1}, we have:
\begin{equation}
    \mathcal{R}^T > \mathcal{R}^C.
\label{eq:rftc}
\end{equation}
In summary, 
Transformers activate a larger receptive field of input data compared to CNNs, 
which is also supported by empirical research~\cite{raghuVisionTransformersSee2021}.

\paragraph{Model Memorization.}

The concept of model memorization lacks a formal definition in the research field. 
Informally, 
it can be characterized as a label ($\mathbf{y}$) memorization for a specific sample $\mathbf{x}$ if removing this $\mathbf{x}$ from the training set would change the performance of the model predicting $\mathbf{y}$~\cite{feldmanDoesLearningRequire2020,carliniPrivacyOnionEffect2022}.
Other research refers to it as memory capacity, 
assessing how much information a model can store in its parameters~\cite{collinsCapacityTrainabilityRecurrent2017,bubeckNetworkSizeSize2020,rajputExponentialImprovementMemorization2021}.
In either case, 
model memorization is intricately linked to how the model learns from input data. 
The relationship between the receptive field and model memorization can be formulated by:
\begin{equation}
    \text{increase}(\mathcal{M}) \sim \text{increase}(\mathcal{R}),
\label{eq:mem}
\end{equation}
where $\mathcal{R}$ denotes the receptive field and $\mathcal{M}$ denotes the model memorization.
A larger receptive field allows the model to reap more information from the input data, 
and as Transformers inherently possess a larger receptive field, 
they exhibit an enhanced ability to absorb more information.
With~\Cref{eq:rftc,eq:mem}, we have:
\begin{equation}
    \mathcal{M}^T > \mathcal{M}^C,
\label{eq:memtc}
\end{equation}
where $\mathcal{M}^T$ is the model memorization in Transformers, and $\mathcal{M}^C$ is the model memorization in CNNs.

\paragraph{Privacy Leakage.}

Model memorization introduces potential privacy leakage. 
Normally, 
input data $\mathbf{x}$ is considered extractably memorized if an adversary can construct data that makes the model produce $\mathbf{x}$~\cite{nasrScalableExtractionTraining2023}. 
The relationship between model memorization and privacy leakage can be expressed as:
\begin{equation}
    \text{increase}(\mathcal{L}) \sim \text{increase}(\mathcal{M}),
\label{eq:leak}
\end{equation}
where $\mathcal{M}$ is the model memorization and $\mathcal{L}$ denotes the privacy leakage.
An increased capacity for the model to learn from input data correlates with a higher likelihood of privacy leakage.
With~\Cref{eq:memtc,eq:leak}, we have:
\begin{equation}
    \mathcal{L}^T > \mathcal{L}^C,
\end{equation}
where $\mathcal{L}^T$ is the privacy leakage in Transformers, and $\mathcal{L}^C$ is the privacy leakage in CNNs.
In summary, Transformers, employing attention modules and featuring a larger receptive field, inherently enhance model memorization, thereby contributing to increased privacy leakage.

\subsection{The Impact of the Design of Activation Layers}

In~\Cref{sec:impact_of_micro,sec:ablation_study,sec:discussion_micro}, 
we show that removing activation layers could make the model more vulnerable to privacy attacks.
Here we delve into more theoretical discussions.

\paragraph{ReLU vs GELU.}
ReLU and GELU are two widely adopted activation layers in CNNs and Transformers, 
defined as follows:

\begin{align}
    \text{ReLU}(\mathbf{x}) &= \max (0, \mathbf{x}), \\
    \text{GELU}(\mathbf{x}) &= 0.5 \mathbf{x} (1 + \tanh{(\sqrt{\frac{2}{\pi}} (\mathbf{x} + 0.044715 \mathbf{x}^3))}),
\end{align}
where $\mathbf{x}$ is the input of the activation layer.

Both ReLU and GELU introduce non-linearity into the model, 
enabling the learning of complex and hierarchical representations from input data. 
However, 
non-linearity can also bring challenges during model training, 
such as the vanishing gradient and exploding gradient problems, 
making it difficult to update model weights and biases~\cite{leeGELUActivationFunction2023}. 
The ReLU unit may suppress half of its inputs by outputting zero values, 
selectively passing information to the next layer~\cite{luoUnderstandingEffectiveReceptive2016}. 
Moreover, 
the non-differentiability of ReLU at $\mathbf{x} = 0$ poses issues in gradient-based optimization.
The GELU unit, 
designed as a smooth approximation to ReLU, preserves the non-linearity of the model~\cite{leeGELUActivationFunction2023}. 
The adoption of GELU can enhance the model's robustness and generalization, 
making the adversary extract less information from data samples and decrease the attack performance~\cite{baiAreTransformersMore2021}. 
This aligns with our experimental results presented in Step 9 of~\Cref{tab:arch_name_and_result} and~\Cref{fig:act_layer}.

\paragraph{The removal of activation layers.}
As using GELU does not help with improving the privacy attack performance, our analysis shows that removing activation layers and leaving fewer activation layers is the key to increasing the model's vulnerability to privacy attacks (shown in~\Cref{fig:act_layer}).
This is because retaining fewer activation layers allows the model to preserve more information learned from the training data, resulting in increased privacy leakage compared to models with a greater number of activation layers, which can be written as $\mathcal{L}^{\text{fewer act}} > \mathcal{L}^{\text{more act}}$.

\subsection{The Impact of the Design of Stem Layers}

In~\Cref{sec:impact_of_micro,sec:ablation_study,sec:discussion_micro}, we illustrate that the design of stem layers (the "Patchify" step) can result in increased privacy leakage.
Here, we provide more theoretical discussions.

The stem layers in a standard ResNet have a $7 \times 7$ convolution layer with stride $2$.
The "Patchify" strategy introduced by Transformers uses a $4 \times 4$ convolution layer with stride $4$, which is actually a non-overlapping convolution process to make patches of the input data.
Some research~\cite{duWhenConvolutionalFilter2018,goelLearningOneConvolutional2018} theoretically show that non-overlapping patches can easily learn information from input data, comparable to the capabilities of overlapping patches.
Empirical studies~\cite{xiaoEarlyConvolutionsHelp2021} provide further support for the efficacy of this kind of stem layer design with convolutional layers. 
This design is shown to enhance optimization stability and improve the adversary's attack performance, which is demonstrated in our experimental results in~\Cref{sec:impact_of_micro,sec:ablation_study}.
We can formulate this as $\mathcal{L}^{\text{patchify stem}} > \mathcal{L}^{\text{resnet stem}}$.

\subsection{The Impact of the Design of LN Layers}

In~\Cref{sec:impact_of_micro,sec:ablation_study,sec:discussion_micro}, we show that changing BatchNorm to LayerNorm can contribute to increased privacy leakage. 
Here, we provide more theoretical discussions on the implications of this design choice.

Let $\mathbf{x} = (x_1, x_2, ..., x_H)$ and $\mathbf{y}$ be the input of size $H$ and output of a normalization layer.
LayerNorm re-centers and re-scales the input $x$ as
\begin{equation}
    \mathbf{y} = \mathbf{g} \odot \frac{\mathbf{x} - \mu}{\sigma} + \mathbf{b}, \mu = \frac{1}{H} \sum_{i = 1}^H x_i, \sigma = \sqrt{\frac{1}{H} \sum_{i = 1}^H (x_i - \mu)^2 },
\end{equation}
where $\odot$ denotes a dot production operation, $\mu$ and $\sigma$ are the mean and standard deviation of $\mathbf{x}$, bias $\mathbf{b}$ and gain $\mathbf{g}$ are learnable parameters~\cite{baLayerNormalization2016,xuUnderstandingImprovingLayer2019}.

The bias and gain are designed for affine transformation on normalized vectors to enhance performance.
However, these parameters are trained based on the training data and may not adequately consider the input distributions of testing data. 
This increases the risk of overfitting when applying LayerNorm in the model~\cite{xuUnderstandingImprovingLayer2019}.
The overfitting of the model would lead to more privacy leakage of the training data.
As a result, models with LayerNorm could potentially expose sensitive information during privacy attacks, which can be written as $\mathcal{L}^{\text{ln layer}} > \mathcal{L}^{\text{bn layer}}$.

\end{document}